\documentclass[11pt,a4paper]{article}

\usepackage{amssymb}
\usepackage[dvips]{graphicx}
\usepackage{bm}

\usepackage{mathrsfs}
\usepackage{cite}

\begin{document}

\title{The NSVZ relations for ${\cal N}=1$ supersymmetric theories with multiple gauge couplings}

\author{
D.S.Korneev\,${}^a$, D.V.Plotnikov\,${}^b$, K.V.Stepanyantz\,${}^b$, N.A.Tereshina\,${}^b$
\medskip\\
${}^a${\small{\em RWTH Aachen University, Faculty of Mathematics, Computer Science and Natural Science,}}\\
{\small{\em Institute for Theoretical Particle Physics and Cosmology, 52062, Aachen, Germany}}
\medskip
\\
${}^b${\small{\em Moscow State University, Faculty of Physics, Department of Theoretical Physics,}}\\
{\small{\em 119991, Moscow, Russia,}}
}

\maketitle

\begin{abstract}
We investigate the NSVZ relations for ${\cal N}=1$ supersymmetric gauge theories with multiple gauge couplings. As examples, we consider MSSM and the flipped $SU(5)$ model, for which they easily reproduce the results for the two-loop $\beta$-functions. For ${\cal N}=1$ SQCD interacting with the Abelian gauge superfield we demonstrate that the NSVZ-like equation for the Adler $D$-function follows from the NSVZ relations. Also we derive all-loop equations describing how the NSVZ equations for theories with multiple gauge couplings change under finite renormalizations. They allow describing a continuous set of NSVZ schemes in which the exact NSVZ $\beta$-functions are valid for all gauge coupling constants. Very likely, this class includes the HD+MSL scheme, which is obtained if a theory is regularized by Higher covariant Derivatives and divergences are removed by Minimal Subtractions of Logarithms. That is why we also discuss how one can construct the higher derivative regularization for theories with multiple gauge couplings. Presumably, this regularization allows to derive the NSVZ equations for such theories in all loops. In this paper we make the first step of this derivation, namely, the NSVZ equations for theories with multiple gauge couplings are rewritten in a new form which relates the $\beta$-functions to the anomalous dimensions of the quantum gauge superfields, of the Faddeev--Popov ghosts, and of the matter superfields. The equivalence of this new form to the original NSVZ relations follows from the extension of the non-renormalization theorem for the triple gauge-ghost vertices, which is also derived in this paper.
\end{abstract}

\unitlength=1cm

\section{Introduction}
\hspace*{\parindent}

Although supersymmetry has not yet been discovered in direct experiments, there are some indications that physics beyond the Standard model is supersymmetric \cite{Mohapatra:1986uf}. These indications are obtained from the analysis of various quantum corrections, which are compared with experimental data and theoretical considerations. Certainly, the most convincing evidence in favor of supersymmetry is the unification of gauge coupling constants in supersymmetric extensions of the Standard model \cite{Ellis:1990wk,Amaldi:1991cn,Langacker:1991an}. The evolution of couplings is encoded in the $\beta$-functions. To see that the gauge coupling unification becomes much better in the supersymmetric case, it is sufficient to consider the lowest (one-loop) contributions to the $\beta$-functions. However, the higher order corrections are also very important. For ${\cal N}=1$ supersymmetric gauge theories there is an exact expression for the $\beta$-function, which relates it to the anomalous dimension of the matter superfields \cite{Novikov:1983uc,Jones:1983ip,Novikov:1985rd,Shifman:1986zi} and is usually called ``the exact NSVZ $\beta$-function''. As a rule, this equation was analysed for theories with a simple gauge group. This implies that in this case there is the only gauge coupling constant $\alpha$. If Yukawa couplings are denoted by $\lambda$, then the NSVZ equation for a theory with a single gauge coupling constant is written in the form

\begin{equation}\label{Simple_NSVZ}
\beta(\alpha,\lambda) = - \frac{\alpha^2 \Big(3 C_2 - T(R) + C(R)_i{}^j \gamma_j{}^i(\alpha,\lambda)/r\Big)}{2\pi(1-C_2 \alpha/2\pi)},
\end{equation}

\noindent
where $r = \mbox{dim}\, G$ is a dimension of the gauge group,

\begin{equation}
\mbox{tr}(T^A T^B) \equiv T(R) \delta^{AB};\qquad (T^A T^A)_i{}^j \equiv C(R)_i{}^j;\qquad C_2 \equiv T(Adj).
\end{equation}

\noindent
Here $T^A$ denotes the generators of the representation in which the chiral matter superfields lie. They should be distinguished from the generators of the fundamental representation denoted by $t^A$, which are assumed to be normalized by the condition

\begin{equation}\label{Normalization}
\mbox{tr}(t^A t^B) = \frac{1}{2} \delta^{AB}.
\end{equation}

\noindent
From the NSVZ equation we see that the all-loop $\beta$-function for the pure ${\cal N}=1$ supersymmetric Yang--Mills (SYM) theory is given by the geometric series \cite{Jones:1983ip}. Moreover, it is possible to prove \cite{Shifman:1999mv,Buchbinder:2014wra} that the finiteness of ${\cal N}=4$ SYM \cite{Sohnius:1981sn,Grisaru:1982zh,Mandelstam:1982cb,Brink:1982pd,Howe:1983sr} and the finiteness of ${\cal N}=2$ supersymmetric gauge theories beyond the one-loop approximation \cite{Grisaru:1982zh,Howe:1983sr,Buchbinder:1997ib} also follow from Eq. (\ref{Simple_NSVZ}).\footnote{Note that in the latter case a manifestly ${\cal N}=2$ quantization procedure is needed. It can be constructed with the help of the harmonic superspace \cite{Galperin:1984av,Galperin:2001uw,Buchbinder:2001wy} and an invariant regularization \cite{Buchbinder:2015eva}.} Note that due to the ${\cal N}=2$ non-renormalization theorem it is possible to construct ${\cal N}=2$ supersymmetric theories finite in all loops \cite{Howe:1983wj} choosing a gauge group and a representation for the matter superfields in such a way that the one-loop $\beta$-function vanishes.\footnote{Some finiteness preserving terms which break extended supersymmetry have been constructed in \cite{Parkes:1982tg,Parkes:1983ib,Parkes:1983nv}.}  The NSVZ relation can also be applied for investigating ${\cal N}=1$ finite theories \cite{Parkes:1984dh,Kazakov:1986bs,Ermushev:1986cu,Lucchesi:1987he,Lucchesi:1987ef} (see \cite{Heinemeyer:2019vbc} for a recent review). In particular, it allows to prove vanishing of the $(L+1)$-loop $\beta$-function for theories finite in $L$-loops  \cite{Parkes:1985hj,Grisaru:1985tc} in a simple way \cite{Stepanyantz:2021dus}. The exact NSVZ $\beta$-function is also useful for studying fixed points in ${\cal N}=1$ supersymmetric theories, see, e.g., \cite{Seiberg:1994pq,Ryttov:2017khg}. Moreover, in softly broken supersymmetric theories the renormalization of the gaugino masses is described by the NSVZ-like relations \cite{Hisano:1997ua,Jack:1997pa,Avdeev:1997vx}.

It is important that the NSVZ equation is valid only for certain (NSVZ) renormalization prescriptions, which constitute a continuous set \cite{Goriachuk:2018cac,Goriachuk_Conference,Goriachuk:2020wyn}. In particular, it does not hold in the $\overline{\mbox{DR}}$-scheme \cite{Jack:1996vg,Jack:1996cn,Jack:1998uj,Harlander:2006xq,Mihaila:2013wma}, when the theory is regularized by dimensional reduction \cite{Siegel:1979wq} and divergences are removed by modified minimal subtraction \cite{Bardeen:1978yd}. Also it is not valid in the $\mbox{MOM}$-scheme \cite{Kataev:2013csa,Kataev:2014gxa}. Certainly, it is possible to construct an NSVZ scheme by making specially tuned finite renormalizations in each order of the perturbation theory \cite{Jack:1996vg,Jack:1996cn,Jack:1998uj,Harlander:2006xq,Mihaila:2013wma}, see also \cite{Aleshin:2016rrr,Aleshin:2019yqj}. However, there is a simple renormalization prescription giving some NSVZ schemes in all orders. To obtain it, one should regularize a theory by higher covariant derivatives \cite{Slavnov:1971aw,Slavnov:1972sq} in a supersymmetric way \cite{Krivoshchekov:1978xg,West:1985jx}. It is important that this regularization also includes the insertion of the Pauli--Villars determinants for removing one-loop divergences which survive after adding a term with higher derivatives to the action \cite{Slavnov:1977zf}. In the supersymmetric case such a construction can be found in Refs. \cite{Aleshin:2016yvj,Kazantsev:2017fdc}. With the Higher covariant Derivative regularization an NSVZ scheme can be obtained if divergences are removed by Minimal Subtractions of Logarithms or, in other words, in the HD+MSL scheme. In this scheme only powers of $\ln\Lambda/\mu$ are included into renormalization constants, where $\Lambda$ is the dimensionful cut-off parameter of the regularized theory and $\mu$ is the renormalization point \cite{Shakhmanov:2017wji,Stepanyantz:2017sqg}. Equivalently, the HD+MSL scheme can be introduced by imposing certain boundary conditions on renormalization constants \cite{Kataev:2013eta}. Note that minimal subtractions of logarithms can supplement various versions of the higher derivative regularization (which differ in the form of the higher derivative terms and in the Pauli--Villars masses), so that the HD+MSL prescription in general produces a certain set of NSVZ schemes.

The all-loop proof that the HD+MSL prescription gives some NSVZ schemes \cite{Stepanyantz:2020uke} is based on the all-loop derivation of the NSVZ equation in Refs. \cite{Stepanyantz:2016gtk,Stepanyantz:2019ihw,Stepanyantz:2020uke}, see also Ref. \cite{Stepanyantz:2019lfm}.\footnote{Earlier a similar derivation has been made for ${\cal N}=1$ SQED \cite{Stepanyantz:2011jy}, see also \cite{Stepanyantz:2014ima}. It allowed proving that in this case NSVZ schemes are obtained under the HD+MSL \cite{Kataev:2013eta} and on-shell \cite{Kataev:2019olb} prescriptions. In the corresponding softly broken theory the NSVZ-like equation for the renormalization of the photino mass also holds in the HD+MSL scheme \cite{Nartsev:2016nym,Nartsev:2016mvn}.} The NSVZ equation is obtained as follows:

1. According to \cite{Stepanyantz:2016gtk} the triple gauge-ghost vertices (with two ghost external lines and one external line of the quantum gauge superfield) are finite in all orders. (This statement has been verified in the one- and two-loop approximations by explicit calculations made in \cite{Stepanyantz:2016gtk,Kuzmichev:2021yjo}.)

2. Using the finiteness of the triple gauge-ghost vertices the NSVZ $\beta$-function can rewritten in an equivalent form \cite{Stepanyantz:2016gtk}, which relates the $\beta$-function to the anomalous dimensions of the quantum gauge superfield, of the Faddeev-Popov ghosts, and of the matter superfields,

\begin{equation}\label{Simple_NSVZ_New}
\frac{\beta(\alpha,\lambda)}{\alpha^2} = -\frac{1}{2\pi} \Big(3C_2 - T(R) - 2C_2\gamma_c(\alpha,\lambda) - 2C_2\gamma_V(\alpha,\lambda) + \frac{1}{r} C(R)_i{}^j \gamma_j{}^i(\alpha,\lambda) \Big).
\end{equation}

3. In \cite{Stepanyantz:2019ihw} it was proved that with the higher covariant derivative regularization the integrals giving the $\beta$-function are integrals of double total derivatives in the momentum space.\footnote{This fact was first noted in calculating the lowest quantum corrections in \cite{Soloshenko:2003nc} (the factorization into total derivatives) and \cite{Smilga:2004zr} (the factorization into double total derivatives). Subsequently, it has been confirmed by numerous multiloop calculations for various supersymmetric theories, see, e.g., \cite{Pimenov:2009hv,Stepanyantz:2011cpt,Stepanyantz:2011bz,Stepanyantz:2012zz,Kazantsev:2014yna,Shakhmanov:2017soc,Kazantsev:2018nbl}. It should also be noted that in the case of using dimensional reduction the structure of loop integrals is different \cite{Aleshin:2015qqc}.}

4. The integrals of double total derivatives are nontrivial due to singularities of the integrands, which appear due to the identity

\begin{equation}
\frac{\partial^2}{\partial Q^\mu \partial Q_\mu} \Big(\frac{1}{Q^2}\Big) = -4\pi^2 \delta^4(Q).
\end{equation}

\noindent
The sum of these singularities has been calculated in \cite{Stepanyantz:2020uke}. (The sums of ghost and matter singularities have also been calculated by a different method in Ref. \cite{Stepanyantz:2019lfm}.) As the result, the NSVZ equation (\ref{Simple_NSVZ_New}) is obtained for the renormalization group functions (RGFs) defined in terms of the bare couplings for an {\it arbitrary} renormalization prescription supplementing the higher covariant derivative regularization.

5. For RGFs defined in terms of the renormalized couplings some NSVZ schemes are obtained using the fact that in the HD+MSL scheme RGFs defined in terms of the bare couplings coincide with the ones defined in terms of the renormalized couplings up to a formal change of the argument \cite{Kataev:2013eta}.

Although the NSVZ equation was usually studied for theories with a single gauge coupling constant, its analogs can be written for theories with multiple gauge couplings (in which gauge groups are given by direct products of simple and/or $U(1)$ factors) \cite{Ghilencea:1999cy}. For example, the all-loop equations describing the renormalization of the Minimal Supersymmetric Standard Model (MSSM) couplings written in \cite{Shifman:1996iy} are actually the integrated form of the NSVZ relations in this case. It was verified \cite{Ghilencea:1997mu} that they really agree with the explicit two-loop calculations. The NSVZ and NSVZ-like relations were applied in MSSM and its extensions for investigating various quantum corrections \cite{Jack:2004ch,Jack:2005id}, although for obtaining the higher-loop results in the $\overline{\mbox{DR}}$-scheme a proper tuning of a subtraction scheme was certainly needed. Also the NSVZ equations were used for studying the reduction of couplings in theories with multiple gauge couplings, see \cite{Heinemeyer:2019vbc,Mondragon:2013aea,Heinemeyer:2014vxa} and references therein. In Refs. \cite{Shifman:2014cya,Shifman:2015doa} the equation similar to the Abelian NSVZ relation \cite{Vainshtein:1986ja,Shifman:1985fi} was written for the Adler $D$-function \cite{Adler:1974gd} in ${\cal N}=1$ supersymmetric quantum chromodynamics (SQCD) interacting with the Abelian gauge superfield. (The theory considered in \cite{Shifman:2014cya,Shifman:2015doa} has the gauge group $SU(N)\times U(1)$ and, therefore, two gauge coupling constants.) Thus, the NSVZ equations are important for phenomenologically interesting theories with multiple gauge couplings and deserve a more detailed consideration in this case. In particular, it is necessary to derive them using the methods of the perturbation theory and specify renormalization prescriptions under which they are valid. In this paper we start this investigation considering as examples MSSM (see, e.g., \cite{Mohapatra:1986uf}) and the flipped $SU(5)$ Grand Unification Theory (GUT) \cite{Barr:1981qv,Antoniadis:1987dx,Campbell:1987eb,Ellis:1988tx}.

The paper is organized as follows. In Sect.~\ref{Section_NSVZ} we recall a form of the NSVZ relations for theories with multiple gauge couplings. In the next Sect.~\ref{Section_Examples} these equations are written for ${\cal N}=1$ SQCD interacting with the ${\cal N}=1$ supersymmetric Abelian gauge superfield, MSSM, and the flipped $SU(5)$ model. The scheme dependence of the NSVZ relations for theories with multiple gauge couplings is investigated in Sect.~\ref{Section_Schemes}. We present equations describing how they change under finite renormalizations of couplings and matter superfields, and construct finite renormalizations which transfer an NSVZ scheme into another NSVZ scheme. In particular, this implies that the NSVZ schemes constitute a continuous sets. In theories with a single gauge coupling some schemes of this class in all loops are obtained with the help of the HD+MSL prescription. That is why in Sect.~\ref{Section_HD_Regularization} we explain how one can regularize theories with multiple gauge couplings by higher covariant derivatives. The general construction is illustrated by the examples of MSSM and the flipped $SU(5)$ model. This regularization can be used a starting point for the all-loop derivation of the NSVZ equations for the theories under consideration. In theories with a single gauge coupling such a derivation produces a new form of the NSVZ equation (\ref{Simple_NSVZ_New}). Its generalization to the case of multiple gauge couplings is constructed in Sect.~\ref{Section_Triple_Vertices_And_New_Beta} using the extension of the non-renormalization theorem for the triple gauge-ghost vertices, which is also derived in this section.

\section{The NSVZ equations for theories with multiple gauge couplings}
\hspace*{\parindent}\label{Section_NSVZ}

In this paper we will consider a general renormalizable ${\cal N}=1$ supersymmetric theory with the gauge group

\begin{equation}\label{Gauge_Group}
G = G_1\times G_2 \times \ldots \times G_n,
\end{equation}

\noindent
where for each $K = 1,\ldots, n$ the subgroup $G_K$ is either a simple compact group or $U(1)$. The chiral matter superfields can be split into sets (numerated by the index $\mbox{a}$) in such a way that each of these sets transforms under certain irreducible representations $R_{\mbox{\scriptsize a} K}$ of the simple subgroups $G_K$ or has certain charges $q_{\mbox{\scriptsize a} K}$ with respect to $G_K = U(1)$. This implies that for a fixed $\mbox{a}$ the matter superfields are also numerated by indices corresponding to various subgroups $G_K$. It is convenient to introduce the index $i$ numerating all matter superfields $\phi_i$ as the set

\begin{equation}
i = \{\mbox{a};\, i_1, i_2, \ldots, i_n\} \equiv \{\mbox{a};\, i_{\mbox{\scriptsize a}}\}.
\end{equation}

\noindent
It is important that for different values of $\mbox{a}$ the sets of $i_1,\ldots i_n$ values are in general different.

In this notation the generators of the gauge group $G$ corresponding to the subgroup $G_K$ can be written as

\begin{equation}
\left(T^{A_K}\right)_i{}^j = \delta_{\mbox{\scriptsize a}}{}^{\mbox{\scriptsize b}}\cdot \delta_{i_1}{}^{j_1}\ldots \delta_{i_{K-1}}{}^{j_{K-1}} \left(T_{\mbox{\scriptsize a}}^{A_K}\right)_{i_K}{}^{j_K} \delta_{i_{K+1}}{}^{j_{K+1}}\ldots \delta_{i_n}{}^{j_n},
\end{equation}

\noindent
where $\left(T_{\mbox{\scriptsize a}}^{A_K}\right)_{i_K}{}^{j_K}$ are either the generators of $G_K$ in the representation $R_{\mbox{\scriptsize a} K}$ for simple subgroups $G_K$ or the charges of the superfields
$\phi_i \equiv \phi_{\mbox{\scriptsize a};\, i_1 i_2 \ldots i_n}$ with respect to  $G_K = U(1)$. Note that we always assume that the generators are normalized by the conditions

\begin{eqnarray}
&& \mbox{tr} \left(t^{A_K} t^{B_K}\right) = \frac{1}{2} \delta^{A_K B_K} \qquad\ \ \mbox{for simple subgroups;}\quad\\
&& \left(T^{A_K}\right)_{i_K}{}^{j_K} \to q_K\qquad\qquad\quad \mbox{for $U(1)$ subgroups},
\end{eqnarray}

\noindent
where $t^{A_K}$ are generators of the fundamental representation of a simple subgroup $G_K$.

${\cal N}=1$ supersymmetric theories with the gauge group (\ref{Gauge_Group}) in the massless limit are described by the action

\begin{eqnarray}\label{Classical_Action}
&& S = \sum\limits_{K=1}^n  \mbox{Re}\, \frac{1}{4}\int d^4x\, d^2\theta\, \left(W^a\right)^{A_K} \left(W_a\right)^{A_K} + \frac{1}{4} \int d^4x\, d^4\theta\, \phi^{*i} \big(e^{2V}\big)_{i}{}^{j} \phi_j\qquad\nonumber\\
&& +\Big(\frac{1}{6}\lambda_0^{ijk} \int d^4x\, d^2\theta\, \phi_i \phi_j \phi_k +\mbox{c.c.}\Big),
\end{eqnarray}

\noindent
which is written in terms of ${\cal N}=1$ superfields. Such a formulation is especially convenient, because in this case ${\cal N}=1$ supersymmetry is a manifest symmetry of the theory. Using a proper regularization and quantization procedure \cite{Gates:1983nr,West:1990tg,Buchbinder:1998qv}, ${\cal N}=1$ supersymmetry can also be made manifest at all steps of calculating quantum corrections.

In Eq. (\ref{Classical_Action}) $V$ denotes the gauge superfield. In the matter part of the action it is given by the expression

\begin{equation}\label{Gauge_Superfield}
V_{i}{}^{j} = \sum\limits_{K=1}^n e_{0K} V^{A_K} \left(T^{A_K}\right)_i{}^j,
\end{equation}

\noindent
where the bare gauge coupling constants are denoted by $e_{0K}$. Below we will also use the notation

\begin{equation}
V_K \equiv \left\{\begin{array}{l}
e_{0K} V^{A_K} t^{A_K} \qquad (\mbox{or}\ e_{0K} V^{A_K} T^{A_K}) \qquad \mbox{if $G_K$ is non-Abelian};\\
\vphantom{1}\\
V^{A_K} \qquad\qquad\qquad\qquad\qquad\qquad\qquad\quad \mbox{if $G_K = U(1)$}.
\end{array}
\right.
\end{equation}

\noindent
The gauge superfield strengths corresponding to various subgroups $G_K$ in the direct product (\ref{Gauge_Group}) are defined as

\begin{eqnarray}
&& e_{0K} \left(W_a\right)^{A_K} \left(t^{A_K}\right)_{i_k}{}^{j_k} \equiv \frac{1}{8} \bar D^2\Big[\exp\left(-2 e_{0K} V^{A_K} t^{A_K}\right)  D_a \exp\left( 2 e_{0K} V^{A_K} t^{A_K}\right) \Big]_{i_K}{}^{j_K}\qquad
\nonumber\\
&&\qquad\qquad\qquad\qquad\qquad\qquad\qquad\qquad\qquad\qquad\qquad\qquad\qquad \mbox{for simple subgroups;}\qquad\\
&& \left(W_a\right)^{A_K} \equiv \frac{1}{4} \bar D^2 D_a V^{A_K}\qquad \mbox{for $U(1)$ subgroups}.
\end{eqnarray}

All expressions $\left(W^a\right)^{A_K} \left(W_a\right)^{A_K}$ are invariant under gauge transformations of the whole group $G$, so that there are $n$ invariants of this structure. Therefore, the theory (\ref{Classical_Action}) has $n$ gauge coupling constants $\alpha_K = e_K^2/4\pi$, each of them corresponding to a certain subgroup of the group $G$. In terms of the renormalized couplings the corresponding $n$ $\beta$-functions are defined by the equations

\begin{equation}\label{Beta_Definition}
\beta_K(\alpha,\lambda) \equiv \left.\frac{d\alpha_K}{d\ln\mu}\right|_{\alpha_0,\lambda_0 = \mbox{\scriptsize const}},
\end{equation}

\noindent
where the subscript $0$ marks the bare gauge and Yukawa couplings, and $\mu$ stands for the renormalization point.

Taking into account that chiral matter superfields $\phi_{\mbox{\scriptsize a}}$ belong to irreducible representations of all simple subgroups $G_K$, the renormalization constants for them have no indices,

\begin{equation}
(\phi_{\mbox{\scriptsize a}, R})_{i_1 i_2\ldots i_K} = (Z_{\mbox{\scriptsize a}})^{1/2} (\phi_{\mbox{\scriptsize a}})_{i_1 i_2\ldots i_K},
\end{equation}

\noindent
where the subscript $R$ denotes the renormalized superfields. In terms of the renormalized coupling constant the corresponding anomalous dimensions are defined by the equation

\begin{equation}\label{Gamma_Definition}
\gamma_{\mbox{\scriptsize a}}(\alpha,\lambda) \equiv \frac{d\ln Z_{\mbox{\scriptsize a}}}{d\ln\mu}\bigg|_{\alpha_0,\lambda_0 = \mbox{\scriptsize const}}.
\end{equation}

Earlier we denoted the generators of the representation $R_{\mbox{\scriptsize a}K}$ by $T_{\mbox{\scriptsize a}}^{A_K}$. This implies that

\begin{equation}\label{C(R)_Definition}
(T_{\mbox{\scriptsize a}}^{A_K} T_{\mbox{\scriptsize a}}^{A_K})_{i_K}{}^{j_K} = C(R_{\mbox{\scriptsize a} K})\, \delta_{i_K}{}^{j_K}.
\end{equation}

\noindent
It is important that the right hand side of this equation is proportional to the $\delta$-symbol because, by construction, the representation $R_{\mbox{\scriptsize a}K}$ is irreducible. (If $G_K=U(1)$, then $T_{\mbox{\scriptsize a}}^{A_K} \to q_{\mbox{\scriptsize a}K}$ are numbers.) Also it is useful to define the analog of $T(R)$ with the help of the equation

\begin{equation}\label{T_K_Small}
T_K(R_{\mbox{\scriptsize a} K})\, \delta^{{A_K}{B_K}} \equiv (T_{\mbox{\scriptsize a}}^{A_K} T_{\mbox{\scriptsize a}}^{B_K})_{i_K}{}^{i_K}.
\end{equation}

\noindent
Then the matter contribution to the one-loop $\beta$-function corresponding to the coupling constant $\alpha_K$ will be proportional to

\begin{equation}\label{T_K_Bold}
\bm{T}_K(R) = \sum\limits_{\mbox{\scriptsize a}} \bm{T}_{\mbox{\scriptsize a}K},
\end{equation}

\noindent
where we introduced the notation

\begin{equation}\label{T_aK}
\bm{T}_{\mbox{\scriptsize a}K} = \left\{
\begin{array}{l}
{\displaystyle \delta_{i_1}{}^{i_1}\ldots \delta_{i_{K-1}}{}^{i_{K-1}} T_K(R_{\mbox{\scriptsize a} K})\, \delta_{i_{K+1}}{}^{i_{K+1}}\ldots \delta_{i_n}{}^{i_n}\qquad \mbox{if}\ G_K\ \mbox{is simple};}\\
\\
{\displaystyle \delta_{i_1}{}^{i_1}\ldots \delta_{i_{K-1}}{}^{i_{K-1}}\, q_{\mbox{\scriptsize a} K}^2\, \delta_{i_{K+1}}{}^{i_{K+1}}\ldots \delta_{i_n}{}^{i_n}\qquad\qquad\ \mbox{if}\ G_K = U(1).}
\end{array}
\right.
\end{equation}

\noindent
Really, calculating the function $\beta_K$ in the one-loop approximation it is necessary to sum over all indices of chiral superfields corresponding to all subgroups except for $G_K$, while the trace of two generators of the subgroup $G_K$ gives the factor (\ref{T_K_Small}).

It is reasonable to suggest (see, e.g., \cite{Ghilencea:1999cy}) that the exact $\beta$-function for a coupling constant corresponding to a factor $G_K$ in the gauge group $G$ is given by the expression

\begin{equation}\label{NSVZ_Multicharge}
\frac{\beta_K(\alpha,\lambda)}{\alpha_K^2} = - \frac{1}{2\pi(1-C_{2}(G_K) \alpha_K/2\pi)} \Big[\, 3 C_2(G_K) - \sum\limits_{\mbox{\scriptsize a}} \bm{T}_{\mbox{\scriptsize a}K}\Big(1-\gamma_{\mbox{\scriptsize a}}(\alpha,\lambda)\Big) \Big].
\end{equation}

\noindent
(Note that this equation is also valid even if the subgroup $G_K$ coincides with $U(1)$. Certainly, in this case it is necessary to set $C_2(G_K) = 0$.) Really, if $G$ is a simple group, then the representation to which the matter superfields belong can be presented as a direct sum of $R_{\mbox{\scriptsize a}}$, so that

\begin{eqnarray}
&& R = \sum\limits_{\mbox{\scriptsize a}} R_{\mbox{\scriptsize a}};\qquad\ \ \bm{T}_{\mbox{\scriptsize a}K} \to T(R_{\mbox{\scriptsize a}})\equiv \bm{T}_{\mbox{\scriptsize a}};\qquad\ \ T(R) = \sum\limits_{\mbox{\scriptsize a}} T(R_{\mbox{\scriptsize a}}); \nonumber\\
&& \frac{1}{r} C(R)_i{}^j \gamma_j{}^i(\alpha,\lambda) = \frac{1}{r} \sum\limits_{\mbox{\scriptsize a}} C(R_{\mbox{\scriptsize a}}) \gamma_{\mbox{\scriptsize a}}(\alpha,\lambda) \delta_i{}^i = \sum\limits_{\mbox{\scriptsize a}} T(R_{\mbox{\scriptsize a}}) \gamma_{\mbox{\scriptsize a}}(\alpha,\lambda).
\end{eqnarray}

\noindent
Therefore, the NSVZ relation (\ref{Simple_NSVZ}) in this case can be written in the form

\begin{equation}
\frac{\beta(\alpha,\lambda)}{\alpha^2} = - \frac{1}{2\pi(1-C_{2} \alpha/2\pi)} \Big[\, 3 C_2 - \sum\limits_{\mbox{\scriptsize a}} \bm{T}_{\mbox{\scriptsize a}}\Big(1-\gamma_{\mbox{\scriptsize a}}(\alpha,\lambda)\Big) \Big],
\end{equation}

\noindent
which is similar to Eq. (\ref{NSVZ_Multicharge}). Thus, Eq. (\ref{NSVZ_Multicharge}) is a natural generalization of Eq. (\ref{Simple_NSVZ}) to the case of ${\cal N}=1$ supersymmetric theories with multiple gauge couplings.

\section{Examples}
\label{Section_Examples}

\subsection{${\cal N}=1$ SQCD + SQED}
\hspace*{\parindent}\label{Subsection_SQCD_NSVZ}

As a simplest example we consider ${\cal N}=1$ SQCD with $N_f$ flavors interacting with the electromagnetic field in the supersymmetric way. This theory is based on the group $G\times U(1)$ and contains $N_f$ quark flavors in a certain irreducible representation $R$ of the group $G$. (For ``true'' SQCD $G=SU(3)$, $N_f=6$, and $R$ is the fundamental representation.) Each flavor is composed from two chiral matter superfields $\phi_{\mbox{\scriptsize a}}$ and $\widetilde\phi_{\mbox{\scriptsize a}}$ in the representations $R$ and $\bar R$, respectively, with the opposite $U(1)$ charges $q_{\mbox{\scriptsize a}}$ and $-q_{\mbox{\scriptsize a}}$. At the classical level the action in the massless limit is given by the expression

\begin{eqnarray}\label{SQCD+SQED}
&& S = \frac{1}{2g^2}\,\mbox{Re}\,\mbox{tr}\int d^4x\,d^2\theta\, W^a W_a + \frac{1}{4e^2}\,\mbox{Re}\int d^4x\, d^2\theta\, \bm{W}^a \bm{W}_a\nonumber\\
&&\qquad\qquad\qquad\qquad\quad + \sum\limits_{\mbox{\scriptsize a}=1}^{N_f}\, \frac{1}{4}\int d^4x\, d^4\theta\, \Big(\phi_{\mbox{\scriptsize a}}^+ e^{2V + 2q_{\mbox{\scriptsize a}} \bm{V}}\phi_{\mbox{\scriptsize a}}
+ \widetilde\phi_{\mbox{\scriptsize a}}^+ e^{-2V - 2q_{\mbox{\scriptsize a}} \bm{V}} \widetilde\phi_{\mbox{\scriptsize a}}\Big),\qquad
\end{eqnarray}

\noindent
where $V$ and $\bm{V}$ are the gauge superfields corresponding to the subgroups $G$ and $U(1)$, respectively, and the supersymmetric gauge superfield strengths are defined as

\begin{equation}
W_a = \frac{1}{8} \bar D^2\Big(e^{-2V} D_a e^{2V}\Big);\qquad \bm{W}_a = \frac{1}{4} \bar D^2 D_a \bm{V}.
\end{equation}

\noindent
Evidently, there are two coupling constants in the theory. The analog of the strong coupling constant is $\alpha_s\equiv g^2/4\pi$, while the analog of the electromagnetic coupling constant is $\alpha\equiv e^2/4\pi$. We will denote the corresponding $\beta$-functions by $\beta_s(\alpha_s,\alpha)$ and $\beta(\alpha_s,\alpha)$, respectively.

To construct NSVZ expressions for both these $\beta$ functions, we need the values $\bm{T}_{\mbox{\scriptsize a}K}$ defined by Eq. (\ref{T_aK}). It is easy to see that

\begin{equation}
\bm{T}_{\phi_{\mbox{\scriptsize a}}, G} = \bm{T}_{\widetilde\phi_{\mbox{\scriptsize a}}, G} = T(R); \qquad \bm{T}_{\phi_{\mbox{\scriptsize a}}, U(1)} = \bm{T}_{\widetilde\phi_{\mbox{\scriptsize a}}, U(1)} = q_{\mbox{\scriptsize a}}^2\, \mbox{dim}\, R.
\end{equation}

\noindent
Substituting these expressions into Eq. (\ref{NSVZ_Multicharge}) and using the notation $C_2 = C_2(G)$ we obtain the exact expressions for the $\beta$-functions corresponding to the strong and electromagnetic coupling constants,

\begin{eqnarray}
&& \frac{\beta_s(\alpha_s,\alpha)}{\alpha_s^2} = - \frac{1}{2\pi(1-C_{2} \alpha_s/2\pi)} \bigg[\, 3 C_2 - 2 T(R)\sum\limits_{\mbox{\scriptsize a}=1}^{N_f} \Big(1-\gamma_{\mbox{\scriptsize a}}(\alpha_s,\alpha)\Big) \bigg];\qquad\\
&& \frac{\beta(\alpha_s,\alpha)}{\alpha^2} = \frac{1}{\pi}\, \mbox{dim}\,R\, \sum\limits_{\mbox{\scriptsize a}=1}^{N_f} q_{\mbox{\scriptsize a}}^2\Big(1-\gamma_{\mbox{\scriptsize a}}(\alpha_s,\alpha)\Big).
\end{eqnarray}

\noindent
Certainly, they are valid only under certain renormalization prescriptions, which will be discussed below and presumably include the HD+MSL scheme.

Earlier the theory (\ref{SQCD+SQED}) was considered in Refs. \cite{Shifman:2014cya,Shifman:2015doa} (see also \cite{Kataev:2017qvk}), where the corrections to the electromagnetic coupling constant generated by loops of quarks, gluons, and their superpartners were analyzed. These corrections are encoded in the Adler $D$-function \cite{Adler:1974gd}, for which an all-order expression has been obtained. The $D$-function calculated in \cite{Shifman:2014cya,Shifman:2015doa} is very similar to the $\beta$-function for the electromagnetic coupling constant. Up to a normalization factor, the Adler function can be obtained from the function $\beta(\alpha_s,\alpha)/\alpha^2$ by taking the limit $\alpha\to 0$,

\begin{equation}\label{Exact_Adler_Preliminary}
D(\alpha_s) = \frac{3\pi}{2} \lim\limits_{\alpha\to 0} \frac{\beta(\alpha_s,\alpha)}{\alpha^2} =  \frac{3}{2}\, \mbox{dim}\,R \sum\limits_{\mbox{\scriptsize a}=1}^{N_f} q_{\mbox{\scriptsize a}}^2\Big(1-\lim\limits_{\alpha\to 0} \gamma_{\mbox{\scriptsize a}}(\alpha_s,\alpha)\Big).
\end{equation}

\noindent
Note that in the limit $\alpha\to 0$ all chiral matter superfields have the same anomalous dimension,

\begin{equation}\label{General_Gamma}
\lim\limits_{\alpha\to 0} \gamma_{\mbox{\scriptsize a}}(\alpha_s,\alpha) = \gamma(\alpha_s).
\end{equation}

\noindent
Really, the anomalous dimensions of various flavors are different because the electromagnetic charges $q_{\mbox{\scriptsize a}}$ are different for different flavors. In the limit $\alpha\to 0$ it is necessary to omit all superdiagrams containing propagators of the electromagnetic gauge superfield $\bm{V}$, so that the dependence on the electromagnetic charges disappears. Therefore, using Eq. (\ref{General_Gamma}) the exact Adler function (\ref{Exact_Adler_Preliminary}) can be rewritten in the form

\begin{equation}\label{Exact_Adler}
D(\alpha_s) =  \frac{3}{2}\, \mbox{dim}\,R \sum\limits_{\mbox{\scriptsize a}=1}^{N_f} q_{\mbox{\scriptsize a}}^2\Big(1- \gamma(\alpha_s)\Big).
\end{equation}

\noindent
This expression reproduces the result of Refs. \cite{Shifman:2014cya,Shifman:2015doa} in the particular case when $G=SU(N)$ and $R$ is the fundamental representation (so that $\mbox{dim}\, R = N$).\footnote{In the two-loop approximation it gives the expression coincing with the one obtained in Ref. \cite{Kataev:1983at}.} Also Eq. (\ref{Exact_Adler}) agrees with the result of Ref. \cite{Kataev:2017qvk} if we take into account that for an irreducible representation $\gamma_i{}^j = \delta_i^j \cdot \gamma$. A detailed analysis under what renormalization prescriptions Eq. (\ref{Exact_Adler}) is valid has been done in Ref. \cite{Aleshin:2019yqj}. In particular, it holds in the HD+MSL scheme \cite{Kataev:2017qvk}.

\subsection{NSVZ relations for MSSM}
\hspace*{\parindent}\label{Subsection_MSSM_NSVZ}

As another illustration, we consider NSVZ equations for MSSM. This model is a softly broken ${\cal N}=1$ supersymmetric theory with the gauge group

\begin{equation}
G = SU(3) \times SU(2) \times U(1)_Y.
\end{equation}

\noindent
Certainly, this implies that there are three gauge coupling constants

\begin{equation}\label{Couplings_Definitions}
\alpha_3 = \frac{e_3^2}{4\pi};\qquad \alpha_2 = \frac{e_2^2}{4\pi};\qquad  \alpha_1 = \frac{5}{3}\cdot \frac{e_1^2}{4\pi}.
\end{equation}

\noindent
Due to the factor $5/3$ in the definition of $\alpha_1$ the gauge coupling unification condition is written in the simple form $\alpha_1 = \alpha_2 = \alpha_3$. Actually, this factor encodes the value of the Weinberg angle at the scale of Grand Unification, $\sin^2\theta_W = 3/8$.

\begin{table}[h]
\begin{center}
\begin{tabular}{|c||c|c|c|c|c||c|c|}
\hline
$\mbox{a}\vphantom{\Big(}$ & $Q_1,Q_2,Q_3$ & $U_1, U_2, U_3$ & $D_1, D_2, D_3$ & $L_1, L_2, L_3$ & $E_1, E_2, E_3$ &\ \ \ $H_u$\ \ \ &\ \ \ $H_d$\ \ \  \\
\hline
$SU(3)\vphantom{\Big(}$ & $\bar 3$ & 3 & 3 & 1 & 1 & 1 & 1 \\
\hline
$SU(2)\vphantom{\Big(}$ & 2 & 1 & 1 & 2 & 1 & 2 & 2 \\
\hline
$U(1)_Y\vphantom{\Big(}$ & $-1/6$ & $2/3$ & $-1/3$ & $1/2$ & $-1$ & $-1/2$ & $1/2$\\
\hline
\end{tabular}
\end{center}
\caption{Chiral matter superfields in MSSM and their quantum numbers with respect to $SU(3)\times SU(2)\times U(1)$. The subscripts $1$, $2$, and $3$ numerate generations.}\label{Table_MSSM_Quantum_Numbers}
\end{table}

The chiral matter superfields of MSSM include three generations of quarks and leptons and two Higgs douplets

\begin{equation}
H_u = \left(\begin{array}{c} H_{u1}\\ H_{u2} \end{array}\right);\qquad H_d = \left(\begin{array}{c} H_{d1}\\ H_{d2} \end{array}\right).
\end{equation}

\noindent
The model under consideration does not contain right neutrinos. This is not essential for us, because they acquire large masses of the GUT scale order and do not affect running of couplings at low energies. Therefore, each generation includes the chiral superfields $U_a$, $D_a$, $E$,

\begin{equation}
Q^a = \left(\begin{array}{c}\widetilde U^a\\ \widetilde D^a \end{array}\right), \qquad\mbox{and}\qquad L = \left(\begin{array}{c}\widetilde N\\ \widetilde E \end{array}\right).
\end{equation}

\noindent
The right upper quarks, right down quarks, and right charged leptons are components of $U_a$, $D_a$, and $E$, respectively. The superfields $U_a$ and $D_a$ lie in the fundamental representation of $SU(3)$, while $E$-s are $SU(3)$ singlets. All superfields containing right quarks and leptons are $SU(2)$ singlets. The (charge conjugated) left quarks and leptons are components of $Q^a$ and $L$, respectively. These superfields transform under the fundamental representation of $SU(2)$. With respect to $SU(3)$ the superfields $Q^a$ and $L$ belong to the antifundamental and trivial representations, respectively. All chiral matter superfields nontrivially transform under $U(1)$. The corresponding charges (i.e., hypercharges) are presented in Table \ref{Table_MSSM_Quantum_Numbers} together with the $SU(3)$ and $SU(2)$ representation. The subscripts $1$, $2$, and $3$ in this table correspond to the generations. The index $\mbox{a}$ numerates the chiral matter superfields in MSSM that transform under irreducible representations of $SU(3)$ and $SU(2)$.

\vspace{5mm}
\begin{table}[h]
\begin{center}
\begin{tabular}{|c||c|c|c|c|c||c|c|}
\hline
$\mbox{a}\vphantom{\Big(}$ & $Q_1,Q_2,Q_3$ & $U_1, U_2, U_3$ & $D_1, D_2, D_3$ & $L_1, L_2, L_3$ & $E_1, E_2, E_3$ &\ \ \ $H_u$\ \ \ &\ \ \ $H_d$\ \ \ \\
\hline
$SU(3)\vphantom{\Big(}$ & 1 & $1/2$ & $1/2$ & 0 & 0 & 0 & 0 \\
\hline
$SU(2)\vphantom{\Big(}$ & $3/2$ & 0 & 0 & $1/2$ & 0 & $1/2$ & $1/2$ \\
\hline
$U(1)_Y\vphantom{\Big(}$ & 1/6 & 4/3 & 1/3 & 1/2  & 1 & $1/2$ & $1/2$\\
\hline
\end{tabular}
\end{center}
\caption{Values of $\bm{T}_{\mbox{\scriptsize a} K}$ for all MSSM superfields. The index $\mbox{a}$ numerates superfields, and $K = SU(3),\ SU(2),\, U(1)$.}\label{Table_MSSM_T}
\end{table}

Values of $\bm{T}_{\mbox{\scriptsize a} K}$ for all $\mbox{a}$ and $K = SU(3),\ SU(2),\, U(1)$ calculated according to the definition (\ref{T_aK}) are presented in Table \ref{Table_MSSM_T}. Note that for quarks and leptons they correspond to the superfields of a single generation. Using these values it is easy to calculate the (well-known) coefficients which are present in the one-loop MSSM $\beta$-functions,

\begin{eqnarray}
&& 3C_2(SU(3)) - \bm{T}_{SU(3)}(R) = 9 - 3\Big(1+ \frac{1}{2} +\frac{1}{2}\Big) =3;\vphantom{\bigg[}\\
&& 3C_2(SU(2)) - \bm{T}_{SU(2)}(R) = 6 - \bigg[\,3\Big(\frac{3}{2} + \frac{1}{2}\Big) + \frac{1}{2} +\frac{1}{2}\,\bigg] = -1;\qquad\\
&& - \bm{T}_{U(1)}(R) = - \bigg[\,3\Big(\frac{1}{6} +\frac{4}{3} + \frac{1}{3} + \frac{1}{2} + 1\Big) + \frac{1}{2} +\frac{1}{2}\,\bigg] = -11.
\end{eqnarray}

\noindent
However, we are interested in exact all-order expressions for the $\beta$-functions, which can be constructed with the help of Eq. (\ref{NSVZ_Multicharge}). Using the data of Table \ref{Table_MSSM_T} from this equation we conclude that they can be written as

\begin{eqnarray}\label{MSSM_NSVZ_Beta3}
&&\hspace*{-5mm} \frac{\beta_3(\alpha,\lambda)}{\alpha_3^2} = - \frac{1}{2\pi(1 - 3\alpha_3/2\pi)} \bigg[3 + \sum\limits_{I=1}^3\Big(\gamma_{Q_I}(\alpha,\lambda) + \frac{1}{2} \gamma_{U_I}(\alpha,\lambda) + \frac{1}{2} \gamma_{D_I}(\alpha,\lambda)\Big)\bigg];\\
\label{MSSM_NSVZ_Beta2}
&&\hspace*{-5mm} \frac{\beta_2(\alpha,\lambda)}{\alpha_2^2} = - \frac{1}{2\pi(1 - \alpha_2/\pi)} \bigg[-1 + \sum\limits_{I=1}^3\Big(\frac{3}{2} \gamma_{Q_I}(\alpha,\lambda) + \frac{1}{2} \gamma_{L_I}(\alpha,\lambda)\Big) + \frac{1}{2} \gamma_{H_u}(\alpha,\lambda)\nonumber\\
&& + \frac{1}{2} \gamma_{H_d}(\alpha,\lambda)\bigg];\\
\label{MSSM_NSVZ_Beta1}
&&\hspace*{-5mm} \frac{\beta_1(\alpha,\lambda)}{\alpha_1^2} = - \frac{3}{5} \cdot \frac{1}{2\pi}\bigg[-11 + \sum\limits_{I=1}^3\Big(\frac{1}{6} \gamma_{Q_I}(\alpha,\lambda) + \frac{4}{3} \gamma_{U_I}(\alpha,\lambda) + \frac{1}{3} \gamma_{D_I}(\alpha,\lambda) + \frac{1}{2} \gamma_{L_I}(\alpha,\lambda)  \nonumber\\
&&\hspace*{-5mm} + \gamma_{E_I}(\alpha,\lambda)\Big) + \frac{1}{2} \gamma_{H_u}(\alpha,\lambda) + \frac{1}{2} \gamma_{H_d}(\alpha,\lambda)\bigg],
\end{eqnarray}

\noindent
where the index $I$ numerates generations. Note that the factor $3/5$ in the last equation appears due to the factor $5/3$ in the definition of $\alpha_1$, see Eq. (\ref{Couplings_Definitions}). Certainly, Eqs. (\ref{MSSM_NSVZ_Beta3}) --- (\ref{MSSM_NSVZ_Beta1}) are in agreement with the analogous equations written in \cite{Shifman:1996iy} in a different form.

For completeness, in Appendix \ref{Subappendix_Two-Loop_MSSM} we verify that these equations reproduce the correct two-loop expressions for the MSSM $\beta$-functions (which was first made in \cite{Ghilencea:1997mu} starting from the result of Ref. \cite{Shifman:1996iy}). For this purpose in Appendix \ref{Subappendix_Two-Loop_MSSM} we listed the one-loop contributions to the anomalous dimensions entering these equations and substitute them into the NSVZ relations. The result of this calculation exactly coincides with the known expressions for the two-loop $\beta$-functions, see, e.g., \cite{Martin:1993zk}.  Certainly, the coincidence occurs due to the scheme independence of the two-loop contributions to the gauge $\beta$-functions, which will be discussed below in Sect.~\ref{Subsection_Scheme_Dependence}.

\subsection{NSVZ relations for the flipped $SU(5)$ model}
\hspace*{\parindent}\label{Subsection_Flipped_NSVZ}

As another interesting example we consider the so-called flipped $SU(5)$ model \cite{Barr:1981qv,Antoniadis:1987dx,Campbell:1987eb,Ellis:1988tx}. This model is based on the gauge group $SU(5)\times U(1)$ and, therefore, contains two coupling constants. We will denote them by $\alpha_5$ and $\alpha_1$. Certainly, the normalization of $\alpha_1$ should be fixed by a certain condition, which will be described below.

The matter content of the flipped $SU(5)$ theory includes three generations of quarks and leptons which are placed into the representations

\begin{equation}\label{Flipped_Matter_Representations}
3\times \Big(\overline{10} (1) + 5 (-3) + 1 (5)\Big).
\end{equation}

\noindent
In our notation a first symbol denotes an $SU(5)$ representation, while a $U(1)$ charge is written in brackets. Unlike the usual $SU(5)$ model, in the flipped $SU(5)$ model the charged leptons are singlets with respect to $SU(5)$, the representation $\overline{10}$ contains $D$ and $N$ (instead of $U$ and $E$, respectively), and the representation $5$ contains $U$ (instead of $D$),

\begin{equation}
\overline{10}^{ij} \sim \left(
\begin{array}{ccccc}
0 & D_3 & -D_2 & \widetilde U^1 & \widetilde D^1\\
-D_3 & 0 & D_1 & \widetilde U^2 & \widetilde D^2\\
D_2 & -D_1 & 0 & \widetilde U^3 & \widetilde D^3\\
-\widetilde U^1 & -\widetilde U^2 & - \widetilde U^3 & 0 & N\\
-\widetilde D^1 & -\widetilde D^2 & -\widetilde D^3 & -N & 0
\end{array}
\right);\qquad 5_i \sim \left(\begin{array}{c} U_1\\ U_2\\ U_3\\ \widetilde E\\ -\widetilde N
\end{array}\right);\qquad 1\sim E.
\end{equation}

\noindent
The $SU(5)\times U(1)$ symmetry can be broken down to $SU(3)\times SU(2) \times U(1)_Y$ by the vacuum expectation values of two chiral Higgs superfields $H$ and $\widetilde H$ in the representations $10 (-1)$ and $\overline{10} (1)$, respectively. The group $U(1)_Y$ is obtained as a superposition of the $SU(5)$ transformations with

\begin{equation}
\omega_5 = \exp\Big\{\,\frac{i\alpha_Y}{30}\left(
\begin{array}{cc}
2\cdot 1_3 & 0 \\
0 & -3\cdot 1_2
\end{array}
\right)\Big\}
\end{equation}

\noindent
(where $1_3$ and $1_2$ are $3\times 3$ and $2\times 2$ identity matrices, respectively) and the $U(1)$ transformations with $\omega_1 = \exp(-iq\alpha_Y/5)$, where $q$ is a $U(1)$ charge of the corresponding superfield normalized as in Eq. (\ref{Flipped_Matter_Representations}). Also it is necessary to include Higgs superfields responsible for breaking $SU(2)\times U(1)_Y$ down to $U(1)_{\mbox{\scriptsize em}}$. This is done by introducing two chiral superfields $h$ and $\widetilde h$ in the representations $5(2)$ and $\bar 5(-2)$, respectively. Moreover, (for $N_G=3$ generations) the model includes $N_G+1=4$ chiral superfields $\phi_R$ which are singlets with respect to $SU(5)$ and have the $U(1)$ charges equal to 0. The superpotential of the resulting theory is presented in Appendix~\ref{Subappendix_Two-Loop_Flipped}.

Note that the normalization of the $U(1)$ charge adopted in Eq. (\ref{Flipped_Matter_Representations}) is not convenient for physical applications. To construct a more convenient prescription, we recall that the considered model can naturally be embedded into the $SO(10)$ theory, which has a single coupling constant. Therefore, a better requirement for fixing a normalization of $\alpha_1$ is obtained by the requirement that the gauge couplings are unified as $\alpha_1 = \alpha_5$. To find the relation between these two different normalization conditions, we note that the generators of the $SU(5)$ subgroup embedded into $SO(10)$ and normalized by Eq. (\ref{Normalization}) are

\begin{equation}
\frac{i}{\sqrt{2}}
\left(\begin{array}{cc}
0 & t_{\mbox{\scriptsize r}}^A\\
- t_{\mbox{\scriptsize r}}^A & 0
\end{array}
\right) = \frac{1}{\sqrt{2}}\, T(t_{\mbox{\scriptsize r}}^A); \qquad
\frac{1}{\sqrt{2}}
\left(\begin{array}{cc}
t_{\mbox{\scriptsize i}}^A & 0\\
0 & t_{\mbox{\scriptsize i}}^A
\end{array}
\right) = \frac{1}{\sqrt{2}}\, T(t_{\mbox{\scriptsize i}}^A),
\end{equation}

\noindent
In this equation $t_{\mbox{\scriptsize r}}^A$ and $t_{\mbox{\scriptsize i}}^A$ (with $A=1,\ldots, 12$) denote real and purely imaginary generators of the $SU(5)$ fundamental representation, respectively. This implies that $e_5 = e_{10}/\sqrt{2}$. The generator of the $U(1)$ subgroup embedded into $SO(10)$, again, normalized by Eq. (\ref{Normalization}) is written as

\begin{equation}
- \frac{i}{\sqrt{20}}\left(\begin{array}{cc}
0 & -1_5\\
1_5 & 0
\end{array}
\right),
\end{equation}

\noindent
where $1_5$ is the $5\times 5$ identity matrix. The image of this generator in the spinor representation (see, e.g., \cite{Mohapatra:1986uf}) multiplied by $e_{10}$ acting on the $SU(5)$ representations $\overline{10}$, $5$, and $1$ gives
\begin{equation}
\frac{e_{10}}{\sqrt{80}} = \frac{e_5}{\sqrt{40}}, \qquad -\frac{3 e_{10}}{\sqrt{80}} = -\frac{3e_5}{\sqrt{40}}, \qquad \mbox{and} \qquad \frac{5 e_{10}}{\sqrt{80}} = \frac{5e_5}{\sqrt{40}},
\end{equation}

\noindent
respectively. This implies that if normalize $\alpha_1$ by the condition $\alpha_1 = \alpha_5$ for the $SO(10)$ symmetric theory, then the $U(1)$ coupling constant should be defined as

\begin{equation}\label{Flipped_Alpha_1}
\alpha_1 \equiv 40\cdot \frac{e_1^2}{4\pi}
\end{equation}

\noindent
provided the $U(1)$ charges in units of $e_1$ are given by Eq. (\ref{Flipped_Matter_Representations}).

\begin{table}[h]
\begin{center}
\vspace{5mm}
\begin{tabular}{|c||c|c|c||c|c||c|}
\hline
$\mbox{a}\vphantom{\Big(}$ & $\overline{10}_1,\overline{10}_2,\overline{10}_3$ &\ \, $5_1, 5_2, 5_3$\ \, &\  $E_1, E_2, E_3$\ & $\ \ H,\,\widetilde H\ \ $ & $\ \ \ h,\widetilde h\ \ \ $ & $\phi_1,\phi_2,\phi_3,\phi_4$ \\
\hline
$SU(5)\vphantom{\Big(}$ & 3/2 & $1/2$ & $0$ & 3/2  & 1/2 & 0\\
\hline
$U(1)\vphantom{\Big(}$ & 10 & 45 & 25 & 10 & 20 & 0\\
\hline
\end{tabular}
\end{center}
\caption{Values of $\bm{T}_{\mbox{\scriptsize a} K}$ (where $K = SU(5),\ U(1)$) for the chiral matter superfields of the flipped $SU(5)$ model.}\label{Table_Flipped_T}
\end{table}

The coefficients $\bm{T}_{\mbox{\scriptsize a} K}$ for the flipped $SU(5)$ model are presented in Table \ref{Table_Flipped_T}. Note that for the representation $\overline{10}$ the generators are written as

\begin{equation}
\left(T_{\overline{10}}^A\right)^{ij}{}_{kl} = -\frac{1}{2}\Big[\delta^i_k (t^A)_l{}^j - \delta^j_k (t^A)_l{}^i - \delta_l^i (t^A)_k{}^j + \delta^j_l (t^A)_k{}^i\Big],
\end{equation}

\noindent
where $(t^A)_i{}^j$ are the generators of the fundamental representation, so that

\begin{equation}
\mbox{tr}\left(T_{\overline{10}}^A T_{\overline{10}}^B\right) = \left(T_{\overline{10}}^A\right)^{ij}{}_{kl}  \left(T_{\overline{10}}^A\right)^{kl}{}_{ij} = 3\, \mbox{tr}(t^A t^B) = \frac{3}{2} \delta^{AB}.
\end{equation}

\noindent
Certainly, the number $3/2$ here and similar numbers for other representations are well-known (see, e.g., \cite{Slansky:1981yr}), but the explicit calculation is more visual.

Substituting the values $\bm{T}_{\mbox{\scriptsize a} K}$ presented in Table \ref{Table_Flipped_T} into Eq. (\ref{NSVZ_Multicharge}) and taking into account the factor $40$ in Eq. (\ref{Flipped_Alpha_1}) we obtain the exact $\beta$-functions for the flipped $SU(5)$ model in the form

\begin{eqnarray}\label{Flipped_NSVZ_Beta5}
&&\hspace*{-7mm} \frac{\beta_5(\alpha,\lambda)}{\alpha_5^2} =  - \frac{1}{2\pi(1 - 5\alpha_5/2\pi)} \bigg[\, 5 +  \sum\limits_{I=1}^3 \Big(\frac{3}{2} \gamma_{\overline{10}_I}(\alpha,\lambda) +\frac{1}{2} \gamma_{5_I}(\alpha,\lambda)\Big) + \frac{3}{2} \gamma_H(\alpha,\lambda) +\frac{3}{2} \gamma_{\widetilde H}(\alpha,\lambda) \nonumber\\
&&\hspace*{-7mm} + \frac{1}{2} \gamma_h(\alpha,\lambda) + \frac{1}{2} \gamma_{\widetilde h}(\alpha,\lambda)\bigg]. \\
\label{Flipped_NSVZ_Beta1}
&&\hspace*{-7mm} \frac{\beta_1(\alpha,\lambda)}{\alpha_1^2} = \frac{1}{8}\cdot \frac{1}{2\pi}\bigg[\, 60 - \sum\limits_{I=1}^3 \Big(2 \gamma_{\overline{10}_I}(\alpha,\lambda) + 9 \gamma_{5_I}(\alpha,\lambda) +5 \gamma_{E_I}(\alpha,\lambda)\Big)  - 2 \gamma_H(\alpha,\lambda) - 2 \gamma_{\widetilde H}(\alpha,\lambda) \nonumber\\
&&\hspace*{-7mm} - 4 \gamma_h(\alpha,\lambda) - 4 \gamma_{\widetilde h}(\alpha,\lambda)\bigg].
\end{eqnarray}

\noindent
(Certainly, these equations are written for the region with unbroken $SU(5)\times U(1)$ symmetry, i.e., above all thresholds.)

In the one-loop approximation Eqs. (\ref{Flipped_NSVZ_Beta5}) and (\ref{Flipped_NSVZ_Beta1}) exactly agree with the results of Ref. \cite{Ellis:1988tx}. For calculating the two-loop contributions to the $\beta$-function we need to know the one-loop expressions for the anomalous dimensions present in these equations. They are collected in Appendix \ref{Subappendix_Two-Loop_Flipped}. Substituting them into the exact NSVZ relations after some transformations we obtain the two-loop $\beta$-functions, which are also presented in Appendix \ref{Subappendix_Two-Loop_Flipped}. Their parts which do not contain Yukawa couplings agree with the corresponding expressions presented in \cite{Ellis:1988tx}. (Unfortunately, we did not find the expressions for the parts containing the Yukawa couplings for this model.) Therefore, we obtain a nontrivial evidence that the NSVZ equations are also valid for theories with multiple gauge couplings.

\section{NSVZ relations and the ambiguity of choosing a renormalization prescription}
\label{Section_Schemes}

\subsection{Scheme dependence of NSVZ relations for theories with multiple gauge couplings}
\hspace*{\parindent}\label{Subsection_Scheme_Dependence}

It is well known \cite{Jack:1996vg,Jack:1996cn,Jack:1998uj,Harlander:2006xq,Mihaila:2013wma} that the NSVZ relation does not hold for an arbitrary renormalization prescription, and even the most popular $\overline{\mbox{DR}}$ scheme is not NSVZ. The scheme dependence of the NSVZ equation was analysed in \cite{Kutasov:2004xu} for theories with a single (gauge) coupling constant. For theories with simple gauge groups containing Yukawa couplings a similar analysis has been done in \cite{Kataev:2014gxa}. In this paper we extend these results to the case of theories with multiple gauge couplings.

It is known \cite{Vladimirov:1979my} that various renormalization schemes are related by finite renormalizations, which can be written in the form

\begin{equation}\label{Finite_Renormalization}
\alpha_K' = \alpha_K'(\alpha,\lambda); \qquad \lambda'= \lambda'(\alpha,\lambda); \qquad Z_{\mbox{\scriptsize a}}'(\alpha',\lambda',\ln\Lambda/\mu) = z_{\mbox{\scriptsize a}}(\alpha,\lambda) Z_{\mbox{\scriptsize a}}(\alpha,\lambda,\ln\Lambda/\mu),
\end{equation}

\noindent
where $\alpha'(\alpha,\lambda)$, $\lambda'(\alpha,\lambda)$, and $z_{\mbox{\scriptsize a}}(\alpha,\lambda)$ are finite functions. Note that here we will not assume that the finite renormalization of Yukawa couplings is related to the finite renormalization of matter superfields.

Under the finite renormalization (\ref{Finite_Renormalization}) the $\beta$-functions change as \cite{Vladimirov:1975mx,Vladimirov:1979ib}

\begin{equation}\label{New_Beta}
\beta_K'(\alpha',\lambda') = \frac{\partial \alpha_K'}{\partial \alpha_L} \beta_L(\alpha,\lambda) + \frac{\partial \alpha'_K}{\partial\lambda^{ijk}}\, (\beta_\lambda)^{ijk}(\alpha,\lambda) + \frac{\partial \alpha'_K}{\partial\lambda^*_{ijk}}\, (\beta_\lambda^*)_{ijk}(\alpha,\lambda),
\end{equation}

\noindent
where the summation over the index $L$ is performed from 1 to $n$ (see Eq. (\ref{Gauge_Group})), and the Yukawa $\beta$-function is defined as

\begin{equation}
(\beta_\lambda)^{ijk}(\alpha,\lambda) \equiv \frac{d\lambda^{ijk}}{d\ln\mu}\bigg|_{\alpha_0,\lambda_0 = \mbox{\scriptsize const}}.
\end{equation}

\noindent
For theories with a single gauge coupling the two-loop $\beta$-function is scheme independent \cite{Vladimirov:1979ib}. For multicharge theories containing Yukawa couplings the Yukawa $\beta$-functions are scheme-dependent starting from the two-loop approximation. However, the gauge $\beta$-functions even for theories with multiple gauge couplings are scheme-independent in the two-loop approximation \cite{Martin:1993zk}. To prove this statement carefully, we should take into account that finite renormalizations should have the same structure as the corresponding quantum corrections \cite{Jack:1996vg,Kataev:2013csa,Jack:2016tpp}. In particular, this implies that the new gauge couplings $\alpha'_K$ are related to the original ones by the equations

\begin{equation}
\frac{1}{\alpha_K'} = \frac{1}{\alpha_K} + C_K + O(\alpha,\lambda^2),
\end{equation}

\noindent
where $C_K$ are some constants. Consequently, the lowest term in the expression for $\alpha_K'$ contains only $\alpha_K^2$,

\begin{equation}
\alpha_K' = \alpha_K - C_K \alpha_K^2 + O(\alpha^3,\alpha^2\lambda^2).
\end{equation}

\noindent
Substituting this expression into Eq. (\ref{New_Beta}) and taking into account that in the one-loop approximation $\beta_K \sim \alpha_K^2$ and that the one-loop expression for $(\beta_\lambda)^{ijk}$ is given by a sum of terms proportional to $\alpha\lambda$ and $\lambda^3$ we obtain

\begin{eqnarray}\label{Beta_Prime_Auxiliary}
&& \beta_K'\Big(\alpha_K - C_K \alpha_K^2 + O(\alpha^3,\alpha^2\lambda^2),\, \lambda + O(\alpha\lambda,\lambda^3)\Big)\nonumber\\
&& \qquad\qquad\qquad\qquad\qquad\qquad = (1-2C_K \alpha_K) \beta_K(\alpha,\lambda) + O(\alpha^4,\alpha^3 \lambda^2,\alpha^2\lambda^4).\qquad
\end{eqnarray}

\noindent
In the lowest approximation each $\beta$-function can be presented in the form

\begin{equation}\label{Beta_Lowest}
\beta_K(\alpha,\lambda) = \frac{\alpha_K^2}{\pi}\beta_{0K} + \beta_{\mbox{\scriptsize 2-loop},K}(\alpha,\lambda) + O(\alpha^4,\alpha^3\lambda^2,\alpha^2\lambda^4),
\end{equation}

\noindent
where $\beta_{0K}$ are numerical constants and $\beta_{\mbox{\scriptsize 2-loop},K}(\alpha,\lambda)$ contains terms proportional to $\alpha^3$ and $\alpha^2\lambda^2$. Substituting the expression (\ref{Beta_Lowest}) into Eq. (\ref{Beta_Prime_Auxiliary}) we see that both sides of the resulting equation have the same dependence of the constants $C_K$,

\begin{eqnarray}
&& \frac{\alpha_K^2}{\pi}\beta_{0K} (1-2C_K \alpha_K) + \beta_{\mbox{\scriptsize 2-loop},K}'(\alpha,\lambda) + O(\alpha^4,\alpha^3\lambda^2,\alpha^2\lambda^4)\nonumber\\
&&\qquad\qquad\qquad\qquad = \frac{\alpha_K^2}{\pi}\beta_{0K} (1-2C_K \alpha_K) + \beta_{\mbox{\scriptsize 2-loop},K}(\alpha,\lambda) + O(\alpha^4,\alpha^3\lambda^2,\alpha^2\lambda^4). \qquad
\end{eqnarray}

\noindent
From this equation we conclude that the two-loop gauge $\beta$-functions are scheme-independent even for theories with multiple gauge couplings,\footnote{If finite renormalizations are not restricted by the perturbative structure of quantum corrections, then the two-loop $\beta$-functions will be scheme-dependent \cite{McKeon:2017mjq}.}

\begin{equation}
\beta_{\mbox{\scriptsize 2-loop},K}'(\alpha,\lambda) = \beta_{\mbox{\scriptsize 2-loop},K}(\alpha,\lambda).
\end{equation}

Under the finite renormalization (\ref{Finite_Renormalization}) the anomalous dimension of chiral matter superfields changes as

\begin{equation}\label{New_Gamma}
\gamma_{\mbox{\scriptsize a}}'(\alpha',\lambda') = \gamma_{\mbox{\scriptsize a}}(\alpha,\lambda) + \frac{\partial\ln z_{\mbox{\scriptsize a}}}{\partial \alpha_L}\, \beta_L(\alpha,\lambda) + \frac{\partial\ln z_{\mbox{\scriptsize a}}}{\partial \lambda^{ijk}}\, (\beta_\lambda)^{ijk}(\alpha,\lambda) + \frac{\partial\ln z_{\mbox{\scriptsize a}}}{\partial \lambda^*_{ijk}}\, (\beta_\lambda^*)_{ijk}(\alpha,\lambda).
\end{equation}

\noindent
Evidently, in the one-loop approximation the anomalous dimension is scheme-independent. This implies that the NSVZ equations (\ref{NSVZ_Multicharge}) do not depend on a renormalization prescription up to the order $O(\alpha,\lambda^2)$ inclusive exactly as for theories with a single gauge coupling constant.

Next, let us construct an equation which describes how the NSVZ equations change under the finite renormalization (\ref{Finite_Renormalization}). For this purpose we first write the transformations inverse to (\ref{New_Beta}) and (\ref{New_Gamma}), which are obtained by making the replacement $\alpha_K' \leftrightarrow \alpha_K$, $\lambda' \leftrightarrow \lambda$, $z_{\mbox{\scriptsize a}} \to (z_{\mbox{\scriptsize a}} )^{-1}$. We will assume that original RGFs (without primes) satisfy the NSVZ equation (\ref{NSVZ_Multicharge}). In this equation we express them in terms of the new RGFs (denoted by primes). This gives the equation

\begin{eqnarray}
&& \frac{\partial \alpha_K}{\partial \alpha'_L} \beta'_L(\alpha',\lambda') + \frac{\partial \alpha_K}{\partial\lambda'{}^{ijk}}\, (\beta'_\lambda)^{ijk}(\alpha',\lambda') + \frac{\partial \alpha_K}{\partial\lambda'{}^*_{ijk}}\, (\beta_\lambda'{}^*)_{ijk}(\alpha',\lambda') = \beta_K(\alpha,\lambda) \nonumber\\
&& = - \frac{\alpha_K^2}{2\pi(1-C_{2}(G_K) \alpha_K/2\pi)} \Big[\, 3 C_2(G_K) - \bm{T}_{K}(R) + \sum\limits_{\mbox{\scriptsize a}} \bm{T}_{\mbox{\scriptsize a}K} \gamma_{\mbox{\scriptsize a}}(\alpha,\lambda) \Big]\nonumber\\
&&  = - \frac{\alpha_K^2}{2\pi(1-C_{2}(G_K) \alpha_K/2\pi)} \Big[\, 3 C_2(G_K) - \bm{T}_{K}(R) + \sum\limits_{\mbox{\scriptsize a}} \bm{T}_{\mbox{\scriptsize a}K}\Big(\gamma'_{\mbox{\scriptsize a}}(\alpha',\lambda')\nonumber\\
&& - \frac{\partial\ln z_{\mbox{\scriptsize a}}}{\partial \alpha'_L}\, \beta'_L(\alpha',\lambda') - \frac{\partial\ln z_{\mbox{\scriptsize a}}}{\partial \lambda'{}^{ijk}}\, (\beta'_\lambda)^{ijk}(\alpha',\lambda') - \frac{\partial\ln z_{\mbox{\scriptsize a}}}{\partial \lambda'{}^*_{ijk}}\, (\beta'_\lambda{}^*)_{ijk}(\alpha',\lambda')\Big)\Big].
\end{eqnarray}

\noindent
Solving it for $\beta_K'(\alpha',\lambda')$ we obtain the equation relating new RGFs,

\begin{eqnarray}\label{NSVZ_After_Transformation}
&& \beta_K'(\alpha',\lambda') =  ({\cal M}^{-1})_K{}^L {\cal P}_L,
\end{eqnarray}

\noindent
where the summation over $L$ is performed from 1 to $n$, and we use the notations

\begin{eqnarray}\label{M_Definition}
&&\hspace*{-11mm} {\cal M}_K{}^L \equiv \Big(1 - \frac{C_2(G_K)\alpha_K}{2\pi}\Big) \frac{\partial\alpha_K}{\partial\alpha_L'} - \frac{\alpha_K^2}{2\pi} \sum\limits_{\mbox{\scriptsize a}} \bm{T}_{\mbox{\scriptsize a}K} \frac{\partial\ln z_{\mbox{\scriptsize a}}}{\partial \alpha'_L};\\
\label{P_Definition}
&&\hspace*{-11mm} {\cal P}_L =  - \frac{\alpha_L^2}{2\pi} \bigg\{ 3 C_2(G_L) - \bm{T}_{L}(R) + \sum\limits_{\mbox{\scriptsize a}} \bm{T}_{\mbox{\scriptsize a}L}\Big(\gamma'_{\mbox{\scriptsize a}}(\alpha',\lambda') - \frac{\partial\ln z_{\mbox{\scriptsize a}}}{\partial \lambda'{}^{ijk}}\, (\beta'_\lambda)^{ijk}(\alpha',\lambda') - \frac{\partial\ln z_{\mbox{\scriptsize a}}}{\partial \lambda'{}^*_{ijk}}\,\nonumber\\
&&\hspace*{-11mm} \times (\beta'_\lambda{}^*)_{ijk}(\alpha',\lambda')\Big)\bigg\}
- \Big(1 - \frac{C_2(G_L)\alpha_L}{2\pi}\Big) \bigg\{\frac{\partial \alpha_L}{\partial\lambda'{}^{ijk}}\, (\beta'_\lambda)^{ijk}(\alpha',\lambda') + \frac{\partial \alpha_L}{\partial\lambda'{}^*_{ijk}}\, (\beta_\lambda'{}^*)_{ijk}(\alpha',\lambda') \bigg\}.
\end{eqnarray}

\noindent
For theories with a single gauge coupling constant this equation agrees with the results of Refs. \cite{Kutasov:2004xu,Kataev:2014gxa}. From Eq. (\ref{NSVZ_After_Transformation}) we see that under arbitrary finite renormalizations the NSVZ equation does not preserve its form. However, as we will see in the next section, there is a certain class of finite renormalizations which do not break the form of the NSVZ relation.

\subsection{A class of the NSVZ schemes}
\hspace*{\parindent}\label{Subsection_Class}

Even for ${\cal N}=1$ SQED there is a continuous set of the NSVZ renormalization schemes \cite{Goriachuk:2018cac}. For non-Abelian theories with a single coupling a similar class of the NSVZ schemes was obtained in \cite{Goriachuk_Conference}. In this section we construct a generalization of these results to the case of theories containing multiple gauge couplings and Yukawa couplings. For this purpose we rewrite Eq. (\ref{NSVZ_Multicharge}) in the equivalent form

\begin{equation}
\frac{\beta_K(\alpha,\lambda)}{\alpha_K^2}\Big(1-C_{2}(G_K) \frac{\alpha_K}{2\pi}\Big) = - \frac{1}{2\pi} \Big[\, 3 C_2(G_K) - \bm{T}_K(R) + \sum\limits_{\mbox{\scriptsize a}} \bm{T}_{\mbox{\scriptsize a}K} \gamma_{\mbox{\scriptsize a}}(\alpha,\lambda) \Big].
\end{equation}

\noindent
Using the definitions of the $\beta$-functions and the anomalous dimensions of matter superfields (Eqs. (\ref{Beta_Definition}) and (\ref{Gamma_Definition}), respectively) this equation can be presented in the form \cite{Shifman:1986zi}

\begin{equation}\label{Original_Mu_Derivative}
\frac{d}{d\ln\mu}\bigg\{\frac{1}{\alpha_K} + \frac{C_{2}(G_K)}{2\pi} \ln \alpha_K  - \frac{1}{2\pi} \Big(3 C_2(G_K) - \bm{T}_K(R)\Big) \ln \frac{\mu}{\Lambda}  - \frac{1}{2\pi} \sum\limits_{\mbox{\scriptsize a}} \bm{T}_{\mbox{\scriptsize a}K} \ln Z_{\mbox{\scriptsize a}} \bigg\} = 0.
\end{equation}

Two different renormlization schemes can always be related by a finite renormalization \cite{Vladimirov:1979my}, which for the considered theory can be written in the form (\ref{Finite_Renormalization}). Let us assume that the NSVZ equations (\ref{NSVZ_Multicharge}) are valid for both original and new couplings. This implies that Eq. (\ref{Original_Mu_Derivative}) also holds for the new couplings. Therefore, they satisfy the equation

\begin{equation}\label{New_Mu_Derivative}
\frac{d}{d\ln\mu}\bigg\{\frac{1}{\alpha_K'} + \frac{C_{2}(G_K)}{2\pi} \ln \alpha_K'  - \frac{1}{2\pi} \Big(3 C_2(G_K) - \bm{T}_K(R)\Big) \ln \frac{\mu}{\Lambda}  - \frac{1}{2\pi} \sum\limits_{\mbox{\scriptsize a}} \bm{T}_{\mbox{\scriptsize a}K} \ln Z'_{\mbox{\scriptsize a}} \bigg\} = 0.
\end{equation}

\noindent
Subtracting Eq. (\ref{Original_Mu_Derivative}) from Eq. (\ref{New_Mu_Derivative}) we obtain the equation which contains only functions which describe the considered finite renormalization,

\begin{equation}
\frac{d}{d\ln\mu}\bigg\{\frac{1}{\alpha_K'} - \frac{1}{\alpha_K} + \frac{C_{2}(G_K)}{2\pi} \ln \frac{\alpha_K'}{\alpha_K} - \frac{1}{2\pi} \sum\limits_{\mbox{\scriptsize a}} \bm{T}_{\mbox{\scriptsize a}K} \ln z_{\mbox{\scriptsize a}} \bigg\} = 0.
\end{equation}

\noindent
We will be interested only in trivial solutions of this equation, which are obtained if the expressions in the curly brackets (numerated by the index $K$) are constants and do not depend on couplings. Denoting them by $B_K$ we derive the system of equations describing finite renormalizations which do not take out of the class of the NSVZ schemes for the considered theory,

\begin{equation}\label{B_Definition}
\frac{1}{\alpha_K'} - \frac{1}{\alpha_K} + \frac{C_{2}(G_K)}{2\pi} \ln \frac{\alpha_K'}{\alpha_K} - \frac{1}{2\pi} \sum\limits_{\mbox{\scriptsize a}} \bm{T}_{\mbox{\scriptsize a}K} \ln z_{\mbox{\scriptsize a}} = B_K.
\end{equation}

\noindent
Evidently, this class is parameterized by $m$ arbitrary functions $z_{\mbox{\scriptsize a}}$ and $n$ arbitrary constants $B_K$, where $m$ is the number of the chiral matter superfields $\phi_{\mbox{\scriptsize a}}$, and $n$ is the number of factors in the product (\ref{Gauge_Group}).

Note that if $B_K$ defined as the left hand side of Eq. (\ref{B_Definition}) are not constants, then the matrices ${\cal M}_K{}^L$ and ${\cal P}_L$ (given by Eqs. (\ref{M_Definition}) and (\ref{P_Definition}), respectively) can be written as

\begin{eqnarray}
&&\hspace*{-9mm} {\cal M}_K{}^L = \Big(\frac{\alpha_K}{\alpha'_K}\Big)^2 \delta_{KL} \Big(1- \frac{C_2(G_K)}{2\pi}\alpha_K'\Big) + \alpha_K^2 \frac{\partial B_K}{\partial \alpha_L'};\\
&&\hspace*{-9mm} {\cal P}_L =  - \frac{\alpha_L^2}{2\pi} \bigg\{ 3 C_2(G_L) - \bm{T}_{L}(R) + \sum\limits_{\mbox{\scriptsize a}} \bm{T}_{\mbox{\scriptsize a}L} \gamma'_{\mbox{\scriptsize a}}(\alpha',\lambda')\bigg\} \nonumber\\
&&\hspace*{-9mm} \qquad\qquad\qquad\qquad\qquad -\alpha_L^2 \bigg\{\frac{\partial B_L}{\partial\lambda'{}^{ijk}}\, (\beta'_\lambda)^{ijk}(\alpha',\lambda') + \frac{\partial B_L}{\partial\lambda'{}^*_{ijk}}\, (\beta_\lambda'{}^*)_{ijk}(\alpha',\lambda')\bigg\}.
\end{eqnarray}

\noindent
From these equations and Eq. (\ref{NSVZ_After_Transformation}) it is evident that the NSVZ equations really remain invariant under such finite renormalizations for that $B_K$ are independent of couplings.

For theories with a single gauge coupling constant a certain subset of NSVZ schemes is produced by the HD+MSL prescription, when a theory is regularized by higher covariant derivatives, and divergences are removed by minimal subtractions of logarithms. Presumably, this is also so for theories with multiple gauge couplings. Note that the HD+MSL prescription is not unique, because minimal subtractions of logarithms can supplement various versions of the higher covariant derivative regularization. The explicit dependence of the two-loop anomalous dimension and of the three-loop $\beta$-function on regularization parameters for a general ${\cal N}=1$ supersymmetric gauge theory with a single gauge coupling in the HD+MSL scheme can be found in Ref. \cite{Kazantsev:2020kfl}. This dependence is not trivial, so that HD+MSL schemes really give a continuous set of NSVZ schemes, which, in general, constitute a certain subset of the whole NSVZ scheme class.

\section{The higher covariant derivative regularization for theories with multiple gauge couplings}
\label{Section_HD_Regularization}

\subsection{General features of the higher derivative regularization in supersymmetric theories with multiple gauge couplings}
\hspace*{\parindent}\label{Subsection_Regularization_General}

For theories with a single gauge coupling constant some NSVZ schemes are given by the HD+MSL prescription in all loops \cite{Stepanyantz:2016gtk,Stepanyantz:2019ihw,Stepanyantz:2020uke}. Possibly, this is also so for theories with multiple gauge couplings. That is why it is expedient to construct the higher covariant derivative regularization for such theories. Very likely, if it is supplemented by minimal subtractions of logarithms, then Eq. (\ref{NSVZ_Multicharge}) is valid in all orders of the perturbation theory.

The versions of the higher covariant derivative regularization which were used for supersymmetric theories in Refs. \cite{Aleshin:2016yvj,Kazantsev:2017fdc} can be introduced only for theories satisfying the anomaly cancellation condition $\mbox{tr}(T^A\{T^B, T^C\}) = 0$ \cite{Kuzmichev:2021yjo}. For theories considered in Sect.~\ref{Section_Examples} this condition really holds, so that the higher covariant derivative regularization can be constructed. Although it is made in a rather standard way, some features specific for the models under consideration should be described in more detail.

To introduce the higher covariant derivative regularization, first, we need to add a higher derivative term to the classical action \cite{Slavnov:1971aw,Slavnov:1972sq}. Next, one should  insert into the generating functional the Pauli--Villars determinants, which cancel one-loop divergences that survived after this \cite{Slavnov:1977zf}. For quantizing the theory we will use the background field method \cite{DeWitt:1965jb,Abbott:1980hw,Abbott:1981ke} in the supersymmetric formulation \cite{Grisaru:1982zh,Gates:1983nr}.  Also it is necessary to take into account that the quantum gauge superfield is renormalized nonlinearly \cite{Piguet:1981fb,Piguet:1981hh,Tyutin:1983rg}, and that the nonlinear renormalization is really needed for cancelling certain divergences \cite{Juer:1982fb,Juer:1982mp} and for the renormalization group equations to be satisfied \cite{Kazantsev:2018kjx}. In principle, it can be reduced to a linear renormalization of an infinite set of parameters (which are very similar to the parameter in the gauge fixing term) included into a certain function ${\cal F}(V)$. Then the background superfield method and the nonlinear renormalization are introduced by making the substitutions

\begin{eqnarray}
&& e^{2V_K} \to e^{2{\cal F}_K(V_K)} e^{2\bm{V}_K}\qquad \mbox{if $G_K$ is non-Abelian}; \vphantom{\Big(}\nonumber\\
&& V_K \to V_K + \bm{V}_K \qquad\qquad\ \, \mbox{if $G_K=U(1)$}, \vphantom{\Big(}
\end{eqnarray}

\noindent
where ${\cal F}_K(V_K) = e_{0K} {\cal F}_K(V_K)^{A_K} T^{A_K}$ and $T^{A_K}$ are the generators of a relevant representation. Note that for Abelian gauge superfields there is no need to introduce the non-linear renormalization, so that in this case ${\cal F}_K(V_K) = V_K$. However, if a subgroup $G_K$ is non-Abelian, then the corresponding function ${\cal F}_K(V_K)$ is not linear. For theories with a single gauge coupling constant its explicit form in the lowest nontrivial approximation can be found in \cite{Juer:1982fb,Juer:1982mp}. Taking into account that gauge superfields corresponding to different subgroups $G_K$ of the gauge group (\ref{Gauge_Group}) commute, the resulting replacement for the gauge superfield (\ref{Gauge_Superfield}) can be written as

\begin{equation}\label{Splitting}
e^{2V} \to e^{2{\cal F}(V)} e^{2\bm{V}}.
\end{equation}

After making this replacement and adding terms containing higher derivatives the gauge part of the action corresponding to a subgroup $G_K$ is modified as

\begin{eqnarray}\label{Modified_Gauge_Part}
&&\hspace*{-7mm} \frac{1}{4}\mbox{Re} \int d^4x\, d^2\theta \left(W^a\right)^{A_K} \left(W_a\right)^{A_K}\ \nonumber\\
&&\hspace*{-7mm}\qquad \to\ \frac{1}{4}\mbox{Re}\, \int d^4x\, d^2\theta \left(W^a\right)^{A_K} \Big[\Big(e^{-2\bm{V}} e^{-2{\cal F}(V)} R\Big(-\frac{\bar\nabla^2 \nabla^2}{16\Lambda^2}\Big) e^{2{\cal F}(V)} e^{2\bm{V}}\Big)_{Adj} W_{a}\Big]^{A_K},
\end{eqnarray}

\noindent
where the gauge superfield strength in the right hand side is defined by the equations

\begin{eqnarray}
&&\hspace*{-7mm} e_{0K} \left(W_{a}\right)^{A_K} t^{A_K} \equiv \frac{1}{8} \bar D^2 \Big(e^{-2\bm{V}_K} e^{-2{\cal F}_K(V_K)} D_a (e^{2{\cal F}_K(V_K)} e^{2\bm{V}_K})\Big)\qquad\mbox{if $G_K$ is non-Abelian};\nonumber\\
&&\hspace*{-7mm} \left(W_a\right)^{A_K} = \frac{1}{4} \bar D^2 D_a V^{A_K} + \frac{1}{4} \bar D^2 D_a \bm{V}^{A_K} \qquad\qquad\qquad\qquad\qquad\quad\ \ \mbox{if\, $G_K=U(1)$.}
\end{eqnarray}

\noindent
(Certainly, if $G_K = U(1)$, the index $A_K$ takes a single value, $V^{A_K} \to V_K$.)

The expression (\ref{Modified_Gauge_Part}) contains the regulator function $R(x)$ rapidly increasing at infinity and satisfying the condition $R(0)$ = 1. For simplicity, we use the same regulator function for all subgroups of the gauge group, although this is of course not necessary. In our notation the covariant derivatives present inside this function are defined as

\begin{equation}
\nabla_a = D_a;\qquad \bar\nabla_{\dot a} = e^{2{\cal F}(V)} e^{2\bm{V}} \bar D_{\dot a} e^{-2\bm{V}} e^{-2 {\cal F}(V)}.
\end{equation}

\noindent
The parameter $\Lambda$ in the argument of the function $R$ has the dimension of mass and plays the role of an ultraviolet cutoff.

Similarly, for all matter superfields it is necessary to make the substitution

\begin{equation}\label{Modified_Matter_Part}
\frac{1}{4} \sum\limits_{\mbox{\scriptsize a}} \int d^4x\, d^4\theta\, \phi_{\mbox{\scriptsize a}}^+ e^{2V} \phi_{\mbox{\scriptsize a}}\ \to\ \frac{1}{4} \sum\limits_{\mbox{\scriptsize a}} \int d^4x\, d^4\theta\, \phi_{\mbox{\scriptsize a}}^+ F\Big(-\frac{\bar\nabla^2 \nabla^2}{16\Lambda^2}\Big) e^{2{\cal F}(V)} e^{2\bm{V}}\phi_{\mbox{\scriptsize a}},
\end{equation}

\noindent
where $F(x)$ is a regulator function which should rapidly grow at infinity and satisfy the condition $F(0)=1$. As in the gauge part of the action, for simplicity, we use the same regulator function for all chiral matter superfields.

In the superpotential we do not introduce any functions with higher derivatives. The regularized action obtained after the modifications described above will be denoted by $S_{\mbox{\scriptsize reg}}$.

The gauge fixing terms should be added for all subgroups $G_K$,

\begin{equation}\label{Gauge_Fixing}
S_{\mbox{\scriptsize gf}} = - \sum\limits_K\, \frac{1}{32\xi_{0K}} \int d^4x\,d^4\theta\, \bm{\nabla}^2 V^{A_K} \Big[K\Big(-\frac{\bm{\bar\nabla}^2 \bm{\nabla}^2}{16\Lambda^2}\Big)_{Adj} \bm{\bar \nabla}^2 V\Big]^{A_K},
\end{equation}

\noindent
where $K(x)$ is another regulator function, $\xi_{0K}$ are bare gauge parameters, and the background covariant derivatives are written as

\begin{equation}
\bm{\nabla}_a = D_a;\qquad \bm{\bar\nabla}_{\dot a} = e^{2\bm{V}} \bar D_{\dot a} e^{-2\bm{V}}.
\end{equation}

For non-Abelian subgroups $G_K$ it is also necessary to introduce the Faddeev--Popov and Nielsen--Kallosh ghosts. Their actions are given by expressions

\begin{eqnarray}
&& S_{\mbox{\scriptsize FP}} = \sum\limits_{K,\, G_K \ne U(1)} \frac{1}{2} \int d^4x\, d^4\theta\, \frac{\partial {\cal F}_K^{-1}(\widetilde V_K)^{A_K}}{\partial \widetilde V^{B_K}}\bigg|_{\widetilde V_K = {\cal F}_K(V_K)}
\Big((e^{2\bm{V}_K})_{Adj} \bar c_K + \bar c^+_K \Big)^{A_K}\nonumber\\
&&\qquad\qquad\qquad\qquad \times \bigg\{\Big(\frac{{\cal F}_K(V_K)}{1-e^{2{\cal F}_K(V_K)}}\Big)_{Adj} c_K^+ + \Big(\frac{{\cal F}_K(V_K)}{1-e^{-2{\cal F}_K(V_K)}}\Big)_{Adj} (e^{2\bm{V}_K})_{Adj} c_K\bigg\}^{B_K};\qquad\\
&& S_{\mbox{\scriptsize NK}} = \sum\limits_{K,\, G_K \ne U(1)} \frac{1}{2 e_{0K}^2} \mbox{tr}\int d^4x\, d^4\theta\, b_K^+ \Big[ K\Big(-\frac{\bm{\bar\nabla}^2 \bm{\nabla}^2}{16\Lambda^2}\Big)\Big]_{Adj} b_K.
\end{eqnarray}

It is important that the replacement $S \to S_{\mbox{\scriptsize reg}}$ removes divergences only beyond the one-loop approximation. A characteristic feature of the higher derivative regularization \cite{Faddeev:1980be} is the presence of residual one-loop divergences, which should be regularized by inserting the Pauli--Villars determinants into the generating functional \cite{Slavnov:1977zf}.

Following \cite{Aleshin:2016yvj,Kazantsev:2017fdc}, to cancel the one-loop (sub)divergences originating from gauge and ghost loops, for each non-Abelian subgroup $G_K$ of the gauge group (\ref{Gauge_Group}) we introduce three (commuting) chiral superfields $\varphi_{1,K}$, $\varphi_{2,K}$, and $\varphi_{3,K}$ in the adjoint representation. Then the one-loop divergences will be regularized if we insert into the generating functional the factor

\begin{eqnarray}
\prod\limits_{K,\,G_K\ne U(1)}\mbox{Det}^{-1}(PV, M_{\varphi,K}) = \int \prod\limits_{K,\, G_K\ne U(1)} D\varphi_{1,K} D\varphi_{2,K} D\varphi_{3,K}\, e^{iS_{\varphi}},
\end{eqnarray}

\noindent
where

\begin{eqnarray}\label{S_Varphi}
&& S_{\varphi} = \frac{1}{2}\sum\limits_{K,\, G_K\ne U(1)} \mbox{tr}\, \bigg\{\int d^4x\, d^4\theta \bigg[ \varphi_{1,K}^+ \Big(R\Big(-\frac{\bar\nabla^2 \nabla^2}{16\Lambda^2}\Big) e^{2{\cal F}(V)} e^{2\bm{V}}\Big)_{Adj} \varphi_{1,K} \nonumber\\
&& + \varphi_{2,K}^+ \Big(e^{2{\cal F}(V)} e^{2\bm{V}}\Big)_{Adj} \varphi_{2,K} + \varphi_{3,K}^+ \Big(e^{2{\cal F}(V)} e^{2\bm{V}}\Big)_{Adj} \varphi_{3,K} \bigg] + \bigg[ \int d^4x\, d^2\theta\, \qquad \nonumber\\
&& \times  M_{\varphi,K} \Big(\varphi_{1,K}^2 + \varphi_{2,K}^2 + \varphi_{3,K}^2 \Big) + \mbox{c.c.}\bigg]\bigg\}.
\end{eqnarray}

\noindent
Note that in this case

\begin{equation}
\Big(e^{2{\cal F}(V)} e^{2\bm{V}}\Big)_{Adj} \varphi_{1,K} = e^{2{\cal F}(V)} e^{2\bm{V}}\varphi_{1,K} e^{-2\bm{V}} e^{-2{\cal F}(V)} = e^{2{\cal F}_K(V_K)} e^{2\bm{V_K}}\varphi_{1,K} e^{-2\bm{V}_K} e^{-2{\cal F}_K(V_K)}.
\end{equation}

\noindent
Similar equations are also valid for the other superfields $\varphi$.

The (sub)divergences generated by a matter loop are regularized by inserting into the generating functional

\begin{equation}\label{PV_Insertion_M}
\prod\limits_{K}\mbox{Det}^{c_K}(PV, M_{K}) = \prod\limits_{K,\, G_K \ne U(1)} \Big(\int D\Phi_K e^{iS_{\Phi,K}}\Big)^{-c_K} \cdot \prod\limits_{K,\, G_K =U(1)} \Big(\int D\Phi_K D\widetilde\Phi_K e^{iS_{\Phi,K}}\Big)^{-c_K},
\end{equation}

\noindent
where $c_K$ are certain numbers which will be specified below (see Eq. (\ref{C_K})). For the non-Abelian subgroups $G_K$ it is possible to choose the commuting chiral Pauli--Villars superfields $\Phi_K$ in the adjoint representation of the subgroup $G_K$. With respect to the other subgroups these superfields are singlets. Then, for regularizing one-loop matter (sub)divergences the actions for them should be taken in the form

\begin{equation}
S_{\Phi,K} = \frac{1}{2}\, \mbox{tr}\, \bigg\{\int d^4x\, d^4\theta\, \Phi_K^+ \Big(F\Big(-\frac{\bar\nabla^2 \nabla^2}{16\Lambda^2}\Big) e^{2{\cal F}(V)} e^{2\bm{V}}\Big)_{Adj} \Phi_K + \bigg[\int d^4x\, d^2\theta\, M_K \Phi_K^2 +\mbox{c.c.} \bigg]\bigg\}.
\end{equation}

\noindent
If a subgroup $G_K$ coincides with $U(1)$, then we introduce two commuting chiral Pauli--Villars superfields $\Phi_K$ and $\widetilde \Phi_K$ with the charges $\pm 1$ with respect to the considered $U(1)$ subgroup. Again, with respect to the other subgroups they are invariant. The action for these superfields is given by the expression

\begin{eqnarray}
&&\hspace*{-5mm} S_{\Phi,K} = \frac{1}{4} \int d^4x\, d^4\theta\, \bigg[\Phi_K^*\, F\Big(-\frac{\bar\nabla^2 \nabla^2}{16\Lambda^2}\Big)  e^{2e_{0K}(V_K+\bm{V}_K)}\Phi_K   \nonumber\\
&&\hspace*{-5mm}\qquad\qquad + \widetilde\Phi_K^*\, F\Big(-\frac{\bar\nabla^2 \nabla^2}{16\Lambda^2}\Big) e^{-2e_{0K}(V_K+\bm{V}_K)}\widetilde\Phi_K \bigg]  + \bigg[\frac{1}{2} \int d^4x\, d^2\theta\, M_K \widetilde\Phi_K \Phi_K +\mbox{c.c.} \bigg].\qquad
\end{eqnarray}

\noindent
The cancellation of the one-loop divergences and subdivergences occurs if the constants $c_K$ in Eq. (\ref{PV_Insertion_M}) have the values

\begin{equation}\label{C_K}
c_K = \left\{\begin{array}{l}
{\displaystyle \frac{\bm{T}_K(R)}{C_2(G_K)}\qquad\, \mbox{for non-Abelian subgroups $G_K$;}}\\
\vphantom{1}\\
{\displaystyle \frac{1}{2}\bm{T}_K(R)\qquad\  \mbox{if $G_K=U(1)$.}}
\end{array}
\right.
\end{equation}

\noindent
(Note that the factor $1/2$ appears because in the Abelian case we introduce two Pauli--Villars superfields.) Also it is important that the masses of the Pauli--Villars superfields should be proportional to the dimensionful parameter $\Lambda$ present in the higher derivative terms,

\begin{equation}
M_{\varphi,K} = a_{\varphi,K} \Lambda;\qquad M_{K} = a_K \Lambda,
\end{equation}

\noindent
the coefficients $a_{\varphi,K}$ and $a_K$ being independent of couplings.

\subsection{The higher covariant derivative regularization for MSSM}
\hspace*{\parindent}\label{Subsection_MSSM_Regularization}

As an example, we construct a version of the higher covariant derivative regularization for MSSM. In this case the regularized action is constructed exactly according to the prescriptions (\ref{Modified_Gauge_Part}) and (\ref{Modified_Matter_Part}) described in the previous section. That is why below we will not discuss this construction in detail. As for the Pauli--Villars determinants, to cancel the one-loop divergences coming from gauge and ghost loops, we insert into the generating functional

\begin{equation}
\mbox{Det}(PV, M_\varphi)^{-1} \equiv \int \prod\limits_{K=SU(3),\, SU(2)} D\varphi_{1,K} D\varphi_{2,K} D\varphi_{3,K} \exp(iS_\varphi).
\end{equation}

\noindent
Here we introduced three commuting Pauli--Villars superfields for both non-Abelian subgroups ($SU(3)$ and $SU(2)$) of the gauge group $SU(3)\times SU(2) \times U(1)$. Evidently, for the $U(1)$ subgroup there is no need to introduce such superfields. According to Eq. (\ref{S_Varphi}) the action for them is written as

\begin{eqnarray}\label{MSSM_Varphi_Action}
&& S_{\varphi} = \frac{1}{2}\sum\limits_{K=SU(3),\, SU(2)} \mbox{tr} \bigg\{\int d^4x\, d^4\theta \bigg[ \varphi_{1,K}^+ \Big(R\Big(-\frac{\bar\nabla^2 \nabla^2}{16\Lambda^2}\Big) e^{2{\cal F}(V)} e^{2\bm{V}}\Big)_{Adj} \varphi_{1,K} \nonumber\\
&& + \varphi_{2,K}^+ \Big(e^{2{\cal F}(V)} e^{2\bm{V}}\Big)_{Adj} \varphi_{2,K} + \varphi_{3,K}^+ \Big(e^{2{\cal F}(V)} e^{2\bm{V}}\Big)_{Adj} \varphi_{3,K} \bigg] + \bigg[ \int d^4x\, d^2\theta\, M_{\varphi,K} \qquad \nonumber\\
&& \times \Big(\varphi_{1,K}^2 + \varphi_{2,K}^2 + \varphi_{3,K}^2 \Big) + \mbox{c.c.}\bigg]\bigg\}.
\end{eqnarray}

\noindent
The corresponding masses are proportional to the parameter $\Lambda$ in the higher derivative terms,

\begin{equation}\label{PV_Varphi_Masses_MSSM}
M_{\varphi,\,SU(3)} \equiv M_{\varphi,3} = a_{\varphi,3} \Lambda;\qquad M_{\varphi,\, SU(2)} \equiv M_{\varphi,2} = a_{\varphi,2} \Lambda,
\end{equation}

\noindent
where the coefficients $a_{\varphi,3}$ and $a_{\varphi,2}$ are constants independent of couplings.

To cancel the one-loop (sub)divergences which appear from a loop of chiral matter superfields, we introduce the chiral Pauli--Villars superfields $\Phi_3$, $\Phi_2$, $\Phi_1$ and $\widetilde\Phi_1$. The superfield $\Phi_3$ lies in the adjoint representation of $SU(3)$ and is invariant under $SU(2)$ and $U(1)$ transformations. Similarly, the superfield $\Phi_2$ lies in the adjoint representation of $SU(2)$ and remains invariant under $SU(3)$ and $U(1)$ transformations. The superfields $\Phi_1$ and $\widetilde \Phi_1$ are singlets with respect to $SU(3)$ and $SU(2)$, but have nontrivial ($\pm 1$ in units of $e_{01}$) opposite $U(1)$ charges. The gauge invariant actions for these superfields have the form

\begin{eqnarray}
&&\hspace*{-5mm} S_{\Phi,3} = \frac{1}{2}\, \mbox{tr}\, \bigg\{\int d^4x\, d^4\theta\, \Phi_3^+ \Big(F\Big(-\frac{\bar\nabla^2 \nabla^2}{16\Lambda^2}\Big) e^{2{\cal F}_3(V_3)} e^{2\bm{V}_3}\Big)_{Adj} \Phi_3 + \bigg[\int d^4x\, d^2\theta\, M_3 \Phi_3^2 +\mbox{c.c.} \bigg]\bigg\};\nonumber\\
&&\hspace*{-5mm} S_{\Phi,2} = \frac{1}{2}\, \mbox{tr}\, \bigg\{\int d^4x\, d^4\theta\, \Phi_2^+ \Big(F\Big(-\frac{\bar\nabla^2 \nabla^2}{16\Lambda^2}\Big) e^{2{\cal F}_2(V_2)} e^{2\bm{V}_2}\Big)_{Adj} \Phi_2 + \bigg[\int d^4x\, d^2\theta\, M_2 \Phi_2^2 +\mbox{c.c.} \bigg]\bigg\};\nonumber\\
&&\hspace*{-5mm} S_{\Phi,1} = \frac{1}{4} \int d^4x\, d^4\theta\, \bigg[\Phi_1^*\, F\Big(-\frac{\bar\nabla^2 \nabla^2}{16\Lambda^2}\Big)  e^{2e_{01}(V_1+\bm{V}_1)}\Phi_1 + \widetilde\Phi_1^*\, F\Big(-\frac{\bar\nabla^2 \nabla^2}{16\Lambda^2}\Big)  e^{-2e_{01}(V_1+\bm{V}_1)}\widetilde\Phi_1 \bigg] \nonumber\\
&&\hspace*{-5mm} + \bigg[\frac{1}{2} \int d^4x\, d^2\theta\, M_1 \widetilde\Phi_1 \Phi_1 +\mbox{c.c.} \bigg],
\end{eqnarray}

\noindent
where

\begin{equation}\label{PV_Phi_Masses_MSSM}
M_1 = a_1 \Lambda; \qquad M_2 = a_2 \Lambda; \qquad M_3 = a_3 \Lambda,
\end{equation}

\noindent
and the coefficients $a_{1,2,3}$ do not depend on couplings. The (inverse) Pauli--Villars determinants are given by the expressions

\begin{eqnarray}
&& \mbox{Det}^{-1}(PV, M_3) = \int D\Phi_3 \exp(iS_{\Phi,3});\qquad\nonumber\\
&& \mbox{Det}^{-1}(PV, M_2) = \int D\Phi_2 \exp(iS_{\Phi,2});\qquad\nonumber\\
&& \mbox{Det}^{-1}(PV, M_1) = \int D\Phi_1 D\widetilde\Phi_1 \exp(iS_{\Phi,1}).
\end{eqnarray}

According to Eq. (\ref{C_K}), to find the degrees of these determinants in the generating functional, we need the coefficients $\bm{T}_{K}(R)$ defined by Eq. (\ref{T_K_Bold}) and the constants $C_2(G_K)$ for the non-Abelian subgroups $G_K$. In the case of MSSM they have the following values:

\begin{eqnarray}
&& \bm{T}_{SU(3)}(R) = 3\Big(1+\frac{1}{2} + \frac{1}{2}\Big) = 6;\qquad\qquad\, C_2(SU(3)) = 3; \quad\\
&& \bm{T}_{SU(2)}(R) = 3\Big(\frac{3}{2} + \frac{1}{2}\Big) + \frac{1}{2} + \frac{1}{2} = 7;\qquad C_2(SU(2)) = 2; \\
&& \bm{T}_{U(1)}(R) = 3\Big(\frac{1}{6} + \frac{4}{3} +\frac{1}{3} + \frac{1}{2} + 1\Big) + \frac{1}{2} + \frac{1}{2} = 11.
\end{eqnarray}

\noindent
Therefore, using Eq. (\ref{C_K}) we conclude that the generating functional for the regularized theory is given by the expression

\begin{eqnarray}\label{MSSM_Z}
&& Z_{\mbox{\scriptsize MSSM}} = \int D\mu\, \mbox{Det}^{2}(PV, M_3)\, \mbox{Det}^{7/2}(PV, M_2)\, \mbox{Det}^{11/2}(PV, M_1)\, \nonumber\\
&&\qquad\qquad\qquad\qquad\qquad\qquad \times \exp\Big(iS_{\mbox{\scriptsize reg}} + i S_{\mbox{\scriptsize gf}} + i S_{\mbox{\scriptsize FP}} + i S_{\mbox{\scriptsize NK}} + i S_\varphi + i S_{\mbox{\scriptsize sources}}), \qquad
\end{eqnarray}

\noindent
where $S_{\mbox{reg}}$ is the regularized action, which in particular includes the higher derivative terms, $S_{\mbox{\scriptsize gf}}$ is the gauge fixing action, $S_{\mbox{\scriptsize FP}}$ and $S_{\mbox{\scriptsize NK}}$ are actions for the Faddeev--Popov and Nielsen--Kallosh ghosts, respectively, $S_\varphi$ is the Pauli--Villars action (\ref{MSSM_Varphi_Action}), and $S_{\mbox{\scriptsize sources}}$ contains all relevant sources.

\subsection{The higher covariant derivative regularization for the flipped $SU(5)$ GUT}
\hspace*{\parindent}\label{Subsection_Flipped_Regularization}

As another example we consider the flipped $SU(5)$ model (for which the gauge group is $SU(5)\times U(1)$). The regularized action is again constructed according to the prescriptions (\ref{Modified_Gauge_Part}) and (\ref{Modified_Matter_Part}) described in Sect. \ref{Subsection_Regularization_General}. The one-loop divergences which remain after this are regularized by inserting the Pauli--Villars determinants.

To cancel the one-loop divergences generated by the gauge and ghost superfields, we use three chiral superfields $\varphi_{1,2,3}$ in the adjoint representation of $SU(5)$ neutral with respect to $U(1)$ and insert into the generating functional the corresponding (inverse) determinant

\begin{equation}
\mbox{Det}^{-1}(PV, M_\varphi) \equiv \int D\varphi_{1} D\varphi_{2} D\varphi_{3} \exp(iS_\varphi),
\end{equation}

\noindent
where

\begin{eqnarray}
&& S_{\varphi} = \frac{1}{2}\, \mbox{tr}\, \bigg\{\int d^4x\, d^4\theta \bigg[\varphi_{1}^+ \Big(R\Big(-\frac{\bar\nabla^2 \nabla^2}{16\Lambda^2}\Big) e^{2{\cal F}(V)} e^{2\bm{V}}\Big)_{Adj} \varphi_{1}  + \varphi_{2}^+ \Big(e^{2{\cal F}(V)} e^{2\bm{V}}\Big)_{Adj} \varphi_{2} \qquad  \nonumber\\
&& + \varphi_{3}^+ \Big(e^{2{\cal F}(V)} e^{2\bm{V}}\Big)_{Adj} \varphi_{3} \bigg] + \bigg[ \int d^4x\, d^2\theta\, M_\varphi\Big(\varphi_{1}^2 + \varphi_{2}^2 + \varphi_{3}^2 \Big) + \mbox{c.c.}\bigg]\bigg\}.
\end{eqnarray}

The divergences coming from a loop of chiral matter superfields are regularized with the help of the (commuting) Pauli--Villars superfields $\Phi_5$, $\Phi_1$, and $\widetilde\Phi_1$. The chiral superfield $\Phi_5$ lies in the adjoint representation of $SU(5)$ and is neutral with respect to the $U(1)$ subgroup. The chiral superfields $\Phi_1$ and $\widetilde \Phi_1$ are $SU(5)$ singlets and have the opposite $U(1)$ charges $\pm e_{01}$, where $e_{01}$ is normalized by Eq. (\ref{Flipped_Matter_Representations}). The actions for these superfields are given by the expressions

\begin{eqnarray}
&&\hspace*{-6mm} S_{\Phi,5} = \frac{1}{2}\, \mbox{tr}\, \bigg\{\int d^4x\, d^4\theta\, \Phi_5^+ \Big(F\Big(-\frac{\bar\nabla^2 \nabla^2}{16\Lambda^2}\Big) e^{2{\cal F}_5(V_5)} e^{2\bm{V}_5}\Big)_{Adj} \Phi_5 + \bigg[\int d^4x\, d^2\theta\, M_5 \Phi_5^2 +\mbox{c.c.} \bigg]\bigg\};\nonumber\\
&&\hspace*{-6mm} S_{\Phi,1} = \frac{1}{4} \int d^4x\, d^4\theta\, \bigg[\Phi_1^*\, F\Big(-\frac{\bar\nabla^2 \nabla^2}{16\Lambda^2}\Big)  e^{2e_{01}(V_1+\bm{V}_1)}\Phi_1 + \widetilde\Phi_1^*\, F\Big(-\frac{\bar\nabla^2 \nabla^2}{16\Lambda^2}\Big)  e^{-2e_{01}(V_1+\bm{V}_1)}\widetilde\Phi_1 \bigg] \nonumber\\
&&\hspace*{-6mm} + \bigg[\frac{1}{2} \int d^4x\, d^2\theta\, M_1 \widetilde\Phi_1 \Phi_1 +\mbox{c.c.} \bigg]
\end{eqnarray}

\noindent
which are used in constructing the corresponding Pauli--Villars determinents

\begin{eqnarray}
&& \mbox{Det}^{-1}(PV, M_5) = \int D\Phi_5 \exp(iS_{\Phi,5});\qquad\nonumber\\
&& \mbox{Det}^{-1}(PV, M_1) = \int D\Phi_1 D\widetilde\Phi_1 \exp(iS_{\Phi,1}).
\end{eqnarray}

\noindent
Standardly, the Pauli--Villars masses are proportional to the parameter $\Lambda$,

\begin{equation}\label{PV_Masses_Flipped}
M_\varphi = a_\varphi \Lambda;\qquad M_1 = a_1 \Lambda; \qquad M_5 = a_5 \Lambda,
\end{equation}

\noindent
and the corresponding ratios are independent of couplings.

To find the degrees of the Pauli--Villars determinants, we need the values

\begin{eqnarray}
&& \bm{T}_{SU(5)}(R) = 3\Big(\frac{3}{2} + \frac{1}{2}\Big) + 2\Big(\frac{3}{2} + \frac{1}{2}\Big) = 10;\qquad C_2(SU(5)) = 5;\\
&& \bm{T}_{U(1)}(R) = 3\Big(10+45+25\Big) + 2\Big(10+20\Big) = 300.
\end{eqnarray}

\noindent
Then, according to the prescription described in Sect.~\ref{Subsection_Regularization_General}, the generating functional for the regularized theory can be written as

\begin{eqnarray}
&& Z_{\mbox{\scriptsize flipped}\, SU(5)} = \int D\mu\, \mbox{Det}^2(PV, M_5)\, \mbox{Det}^{150}(PV, M_1) \nonumber\\
&&\qquad\qquad\qquad\qquad\quad \times \exp\Big(iS_{\mbox{\scriptsize reg}} + i S_{\mbox{\scriptsize gf}} + i S_{\mbox{\scriptsize FP}} + i S_{\mbox{\scriptsize NK}} + i S_\varphi
+ i S_{\mbox{\scriptsize sources}}\Big).\qquad
\end{eqnarray}

\section{Non-renormalization of the triple gauge-ghost vertices and a new form of the NSVZ equation}
\hspace*{\parindent}\label{Section_Triple_Vertices_And_New_Beta}

A very important ingredient of the all-loop derivation of the NSVZ $\beta$-function for theories with a single gauge coupling is a new form of the NSVZ equation \cite{Stepanyantz:2016gtk}. It relates the $\beta$-function in a certain loop to the anomalous dimensions of the quantum gauge superfield, of the Faddeev--Popov ghosts, and of the matter superfields in the previous loop. It is this equation that appears after summing singular contributions \cite{Stepanyantz:2020uke} produced by integrals of double total derivatives which determine the $\beta$-function in the supersymmetric case \cite{Stepanyantz:2019ihw}. In this section we construct such a form of the NSVZ $\beta$-function for theories with multiple gauge couplings. This is done using the non-renormalization theorem for the triple gauge-ghost vertices which is proved for these theories with the help of the Slanov--Taylor identities \cite{Taylor:1971ff,Slavnov:1972fg}. The proof is very similar to the one discussed in \cite{Stepanyantz:2016gtk} for theories with a single gauge coupling, so that here we will describe it more briefly.\footnote{For some theories formulated in terms of usual fields a similar statement is valid in the Landau gauge \cite{Dudal:2002pq,Capri:2014jqa}, see also \cite{Chetyrkin:2004mf} for the verification by an explicit four-loop calculation.} Note that for theories with multiple gauge couplings we will consider only the triple gauge-ghost vertices in which external lines correspond to the same simple subgroup $G_K$. If the external lines correspond to different subgroups, then the divergent contributions are forbidden by the renormalizability.

\subsection{Slavnov--Taylor identities for the triple gauge-ghost vertices in theories with multiple gauge couplings}
\hspace*{\parindent}\label{Subsection_STI}

In the case of using the background superfield method the original gauge invariance of a classical theory generates two types of transformations. The background gauge invariance remains a manifest symmetry of the effective action, while the quantum gauge invariance is broken down to the BRST transformations \cite{Becchi:1974md,Tyutin:1975qk} by the gauge fixing procedure. In the supersymmetric case the BRST transformations are written as \cite{Piguet:1981fb,Piguet:1981mu,Piguet:1984mv}

\begin{eqnarray}\label{BRST_Transformations}
&& \delta V^{A_K} = -\varepsilon_K \frac{\partial {\cal F}_K^{-1}(\widetilde V_K)^{A_K}}{\partial\widetilde V^{B_K}}\bigg|_{\widetilde V_K = {\cal F}_K(V_K)} \bigg\{\bigg(\frac{{\cal F}_K(V_K)}{1-e^{2{\cal F}_K(V_K)}}\bigg)_{Adj} c_K^+\nonumber\\
&&\qquad\qquad\qquad\qquad\qquad\qquad\qquad\qquad + \bigg(\frac{{\cal F}_K(V_K)}{1-e^{-2{\cal F}_K(V_K)}}\bigg)_{Adj} \left(e^{2\bm{V}_K}\right)_{Adj} c_K \bigg\}^{B_K};\qquad\nonumber\\
&& \delta\phi = \sum\limits_K \varepsilon_K c_K \phi \equiv \sum\limits_K \delta_K\phi;\qquad\qquad\quad \delta\bm{V}_K = 0;\qquad\qquad\quad\ \, \delta b_K = 0; \vphantom{\bigg(}\nonumber\\
&& \delta\bar c_K = - \frac{1}{16\xi_{0K}} \varepsilon_K \bar D^2 \Big[\Big(e^{-2\bm{V}_K} K\Big(-\frac{\bm{\nabla}^2 \bm{\bar\nabla}^2}{16\Lambda^2}\Big)\Big)_{Adj} \bm{\nabla}^2 V_K \Big];\qquad\, \delta c_K = \varepsilon_K c_K^2;\qquad\nonumber\\
&& \delta\bar c_K^+ = -\frac{1}{16\xi_{0K}} \varepsilon_K D^2 \Big[ K\Big(-\frac{\bm{\bar\nabla}^2 \bm{\nabla}^2}{16\Lambda^2}\Big)_{Adj} \bm{\bar\nabla}^2 V_K\Big];\qquad\qquad\quad\ \ \
\delta c_K^+ = \varepsilon_K (c_K^+)^2,\qquad
\end{eqnarray}

\noindent
where $\varepsilon_K$ are anticommuting parameters which do not depend on the coordinates. Note that for the gauge and matter superfields the BRST transformations are reduced to the quantum gauge transformations for which the parameters corresponding to non-Abelian subgroups $G_K$ are given by the chiral superfields

\begin{equation}
A_K = \varepsilon_K c_K; \qquad A_K^+ = -\varepsilon_K c_K^+.
\end{equation}

\noindent
This implies that any gauge invariant expression is also BRST invariant.

The BRST invariance leads to the Slavnov--Taylor identities \cite{Taylor:1971ff,Slavnov:1972fg}, which are a key ingredient for proving the finiteness of the triple gauge-ghost vertices. To obtain these identities, we make a change of variables coinciding with the transformations (\ref{BRST_Transformations}) corresponding to the subgroup $G_K$ in the generating functional $Z$. The total action of the theory

\begin{equation}
S_{\mbox{\scriptsize total}} = S_{\mbox{\scriptsize reg}} + S_{\mbox{\scriptsize gf}} + S_{\mbox{\scriptsize FP}} + S_{\mbox{\scriptsize NK}}
\end{equation}

\noindent
is invariant under the transformations (\ref{BRST_Transformations}). By construction, the Pauli--Villars determinants are invariant under both background and quantum gauge transformations. As a consequence of the latter invariance, they are also BRST invariant. Therefore, only the source term

\begin{eqnarray}
&& S_{\mbox{\scriptsize sources}} = \sum\limits_K \bigg\{\int d^4x\, d^4\theta\, V^{A_K} J^{A_K} + \int d^4x\, d^2\theta \Big(j_c^{A_K} c^{A_K} + \bar j_c^{A_K} \bar c^{A_K} \Big)
+  \int d^4x\, d^2\bar\theta \qquad\nonumber\\
&& \times \Big(c^{+A_K} j_c^{+A_K} + \bar c^{+A_K} \bar j_c^{+A_K}\Big) \bigg\} + \sum\limits_{\mbox{\scriptsize a}} \bigg\{ \int d^4x\, d^2\theta\, j^{i_{\mbox{\scriptsize a}}} \phi_{i_{\mbox{\scriptsize a}}} + \int d^4x\, d^2\bar\theta j^*_{i_{\mbox{\scriptsize a}}} \phi^{*i_{\mbox{\scriptsize a}}}\bigg\} \qquad
\end{eqnarray}

\noindent
in the generating functional is not BRST invariant. Using the standard technique (see, e.g., \cite{Faddeev:1980be}) a set of the generating Slavnov--Taylor identities in the considered case can be written as

\begin{eqnarray}\label{Generating_STI}
&&\hspace*{-5mm} \int d^4x\,d^4\theta_x \frac{\delta\Gamma}{\delta V_x^{A_K}} \left\langle \delta V_x^{A_K}\right\rangle + \int d^4x\, d^2\theta_x \bigg(\left\langle \delta\bar c_x^{A_K} \right\rangle \frac{\delta\Gamma}{\delta \bar c_x^{A_K}} + \left\langle \delta c_x^{A_K} \right\rangle \frac{\delta\Gamma}{\delta c_x^{A_K}} + \left\langle \delta_K \phi_{i_{\mbox{\scriptsize a}}, x} \right\rangle
\frac{\delta\Gamma}{\delta \phi_{i_{\mbox{\scriptsize a}}, x}}\bigg)\nonumber\\
&&\hspace*{-5mm} + \int d^4x\, d^2\bar\theta_x \bigg(\left\langle \delta\bar c_x^{+A_K} \right\rangle \frac{\delta\Gamma}{\delta \bar c_x^{+A_K}} + \left\langle \delta c_x^{+A_K} \right\rangle \frac{\delta\Gamma}{\delta c_x^{+A_K}} + \left\langle \delta_K \phi_x^{*i_{\mbox{\scriptsize a}}} \right\rangle \frac{\delta\Gamma}{\delta \phi_x^{* i_{\mbox{\scriptsize a}}}}\bigg) = 0.
\end{eqnarray}

\noindent
Each of these identities corresponds to a certain non-Abelian subgroup of the gauge group. The (super)fields in Eq. (\ref{Generating_STI}) are not set to 0, so that it is possible to differentiate these equalities and construct an infinite set of identities relating various Green functions.

For deriving the non-renormalization theorem we need the identity which is obtained after differentiating Eq. (\ref{Generating_STI}) with respect to $\bar c_y^{+ B_K}$, $c_z^{C_K}$, and $c_w^{D_K}$. After this we set all superfields to 0. In the theories with multiple gauge couplings the ghost number conservations hold for each simple subgroup $G_K$ of the gauge group $G$. Therefore, the numbers of ghosts and antighosts corresponding to any non-Abelian $G_K$ should be equal in each nontrivial Green function. Taking this into account we obtain the Slavnov--Taylor identity

\begin{eqnarray}\label{Triple_STI}
&&\bigg\{ \int d^4x\, d^4\theta_x \bigg(\frac{\delta^3\Gamma}{\delta \bar c^{+B_K}_{y} \delta V_{x}^{A_K} \delta c_{z}^{C_K}} \cdot \frac{\delta}{\delta c_{w}^{D_K}} \left\langle\delta V_{x}^{A_K}\right\rangle -\frac{\delta^3\Gamma}{\delta \bar c^{+B_K}_{y} \delta V_{x}^{A_K} \delta c_{w}^{D_K}} \cdot \frac{\delta}{\delta c_{z}^{C_K}} \left\langle\delta V_{x}^{A_K}\right\rangle\bigg) \qquad\nonumber\\
&& - \int d^4x\, d^2\theta_x \frac{\delta^2\Gamma}{\delta \bar c_y^{+B_K} \delta c_x^{A_K}} \cdot \frac{\delta^2}{\delta c_z^{C_K}\delta c_w^{D_K}} \left\langle \delta c_x^{A_K}\right\rangle\bigg\} \bigg|_{\mbox{\scriptsize fields} = 0} = 0.
\end{eqnarray}

\noindent
(Terms containing matter Green functions vanish due to the global $Z_3$-symmetry $\phi_j \to e^{2\pi i k/3}\phi_j$ of the massless theory.)

The Green functions entering this equation can be expressed in terms of certain functions depending on external momenta. To construct the corresponding expressions, we need to involve chirality and symmetry considerations. For instance, taking into account that the ghost superfields are chiral or antichiral we can write the corresponding two-point Green functions in the form

\begin{equation}\label{Two-Point_Ghost_Function}
\frac{\delta^2 \Gamma}{\delta \bar c_y^{+ B_K} \delta c_x^{A_K}}\bigg|_{\mbox{\scriptsize fields} = 0} = - \frac{D_y^2 \bar D_x^2}{16} G_{c_K} \delta^8_{xy}\, \delta_{A_K B_K}; \qquad
\frac{\delta^2 \Gamma}{\delta \bar c_y^{B_K} \delta c_x^{+A_K}}\bigg|_{\mbox{\scriptsize fields} = 0} = \frac{\bar D_y^2 D_x^2}{16} G_{c_K} \delta^8_{xy}\, \delta_{A_K B_K},
\end{equation}

\noindent
where the functions $G_{c_K}$ depend on couplings and $\partial^2/\Lambda^2$. Evidently, all two-point functions in which ghosts correspond to different subgroups vanish due to the symmetries responsible for the ghost number conservations. Note that in theories with multiple gauge couplings the functions $G_{c_K}$ are in general different for different $K$.

The derivative of $\langle \delta V_x^{A_K}\rangle$ with respect to the ghost superfield can also be expressed in terms of the function $G_{c_K}$. For this purpose we will use the identity obtained with the help of the substitution $\bar c^{+B_K} \to \bar c^{+B_K} + a^{+B_K}$ (where $a^{+B_K}$ is an arbitrary chiral superfield) in the generating functional $Z$. In terms of the effective action the result can be written as the identity

\begin{equation}\label{Ghost_Identity}
\varepsilon_K \frac{\delta\Gamma}{\delta \bar c_x^{+B_K}} = \frac{1}{4} D^2 \left\langle\delta V_x^{B_K} \right\rangle.
\end{equation}

\noindent
Again, in this equation (super)fields are not set to 0, so that it can be differentiated. This implies that Eq. (\ref{Ghost_Identity}) generates an infinite set of identities relating various Green functions. In particular, if we differentiate Eq. (\ref{Ghost_Identity}) with respect to the ghost superfield $c_y^{A_K}$ and take into account chirality consideration, then the result can be written in the form

\begin{equation}\label{Two-Point_Delta_V_Derivative}
\frac{\delta}{\delta c_y^{A_K}} \left\langle \delta V_x^{B_K} \right\rangle\bigg|_{\mbox{\scriptsize fields}=0} = - \varepsilon_K \cdot \frac{1}{4} G_{c_K} \bar D^2 \delta^8_{xy}\, \delta_{A_K B_K}.
\end{equation}

Next, using chirality considerations we write down all possible structures that can appear in the part of the effective action corresponding to the triple $V\bar c^+ c$ vertex,

\begin{eqnarray}\label{S_Definition}
&& \frac{i}{4} \sum\limits_K e_{0K} f^{A_K B_K C_K} \int d^4\theta\, \frac{d^4p}{(2\pi)^4}\, \frac{d^4q}{(2\pi)^4}\, \bar c^{+A_K}(p+q,\theta) \Big[ s_K(p,q)\, \partial^2\Pi_{1/2} V^{B_K}(-p,\theta) \qquad\nonumber\\
&& + {\cal S}_{\mu,K}(p,q)\, (\gamma^\mu)_{\dot a}{}^b D_b \bar D^{\dot a} V^{B_K}(-p,\theta) + {\cal S}_K(p,q)\, V^{B_K}(-p,\theta) \Big] c^{c_K}(-q,\theta),
\end{eqnarray}

\noindent
where $f^{A_K B_K C_K}$ denotes structure constants of the subgroup $G_K$. The functions $s_K$, ${\cal S}_{\mu,K}$, and ${\cal S}_K$ have the dimensions $m^{-2}$, $m^{-1}$, and $m^0$, respectively. Note that in \cite{Stepanyantz:2016gtk} similar functions were denoted by $f$, $F_\mu$, and $F$, but here (following Ref. \cite{Kuzmichev:2021yjo}) we use other letters in order to distinguish them from the higher derivative regulator function $F$, see Eq. (\ref{Modified_Matter_Part}). From Eq. (\ref{S_Definition}) we obtain the explicit expression for the three-point $V\bar c^+ c$ vertex,

\begin{eqnarray}\label{Triple_Vertex}
&&\hspace*{-3mm} \frac{\delta^3\Gamma}{\delta \bar c_x^{+A_K} \delta V_y^{B_K} \delta c_z^{C_K}}\bigg|_{\mbox{\scriptsize fields} = 0} = -\frac{i e_{0K}}{16} f^{A_K B_K C_K} \int \frac{d^4p}{(2\pi)^4} \frac{d^4q}{(2\pi)^4} \Big(s_K(p,q) \partial^2\Pi_{1/2}\nonumber\\
&&\hspace*{-3mm} \qquad\qquad\qquad\qquad\quad - {\cal S}_{\mu,K}(p,q) (\gamma^\mu)_{\dot a}{}^b \bar D^{\dot a} D_b + {\cal S}_K(p,q)\Big)_y \Big(D_x^2 \delta^8_{xy}(q+p)\, \bar D_z^2 \delta^8_{yz}(q) \Big),\qquad
\end{eqnarray}

\noindent
where we introduced the notation

\begin{equation}
\delta_{xy}^8(q)\equiv \delta^4(\theta_x-\theta_y) e ^{iq_\alpha (x^\alpha - y^\alpha)}.
\end{equation}

Also Eq. (\ref{Triple_STI}) contains the second derivative of the expression

\begin{equation}
\left\langle\delta c^{A_K}\right\rangle = \varepsilon_K \cdot \frac{i}{2} e_{0K} f^{A_K B_K C_K} \left\langle c^{B_K} c^{C_K}\right\rangle
\end{equation}

\noindent
with respect to the ghost superfields. It is convenient to introduce the auxiliary source term

\begin{equation}\label{JCC_Vertex}
- \frac{1}{2} \sum\limits_K e_{0K} \int d^4x\, d^2\theta\, f^{A_K B_K C_K} {\cal J}^{A_K} c^{B_K} c^{C_K} + \mbox{c.c.},
\end{equation}

\noindent
which is invariant under the BRST transformations due to their nilpotency. This term allows presenting the considered correlator as the third derivative of the effective action with respect to the source ${\cal J}$ and two ghost superfields. Involving chirality considerations the corresponding contribution to the effective action can be written in the form

\begin{equation}\label{H_Definition}
- \frac{1}{2} \sum\limits_K e_{0K} f^{A_K B_K C_K} \int d^2\theta\, \frac{d^4p}{(2\pi)^4} \frac{d^4q}{(2\pi)^4} c^{A_K}(q+p,\theta) c^{B_K}(-q,\theta) {\cal J}^{C_K} H_K(p,q),
\end{equation}

\noindent
where $H_K(p,q)$ are dimensionless functions. Due to the symmetry with respect to permutations of the ghost superfields they satisfy the equations

\begin{equation}
H_K(p,q) = H_K(p,-q-p).
\end{equation}

\noindent
Then the correlator entering Eq. (\ref{Triple_STI}) can be expressed in terms of these functions,

\begin{eqnarray}\label{J_Correlator}
&& \frac{\delta^2}{\delta c_z^{C_K} \delta c_w^{D_K}} \left\langle\delta c_y^{B_K}\right\rangle\bigg|_{\mbox{\scriptsize fields} = 0} = - i\varepsilon_K \cdot \frac{\delta^3\Gamma}{\delta c_z^{C_K} \delta c_w^{D_K} \delta {\cal J}_y^{B_K}}\bigg|_{\mbox{\scriptsize fields} = 0}\nonumber\\
&&\qquad\qquad\quad = - \frac{ie_{0K}\varepsilon_K}{4} f^{B_K C_K D_K} \int \frac{d^4p}{(2\pi)^4} \frac{d^4q}{(2\pi)^4} H_K(p,q) \bar D_z^2 \delta^8_{zy}(q+p) \bar D_w^2 \delta^8_{yw}(q).\qquad
\end{eqnarray}

Substituting the expressions (\ref{Two-Point_Ghost_Function}), (\ref{Two-Point_Delta_V_Derivative}), (\ref{Triple_Vertex}), and (\ref{J_Correlator}) into the Slavnov--Taylor identity (\ref{Triple_STI}) after some transformations we rewrite this equation in the equivalent form

\begin{equation}\label{STI_Identity_Final}
G_{c_K}(q)\, {\cal S}_K(q,p) + C_{c_K}(p)\, {\cal S}_K(p,q) = 2 G_{c_K}(q+p) H_K(-q-p,q),
\end{equation}

\noindent
where, for simplicity, the argument of the functions $G_c$ is denoted by $q$ instead of $-q^2/\Lambda^2$. We see that for theories with multiple gauge couplings the number of such equations is equal to the number of non-Abelian subgroups $G_K$ in the direct product (\ref{Gauge_Group}).

Eq. (\ref{STI_Identity_Final}) is the most convenient form of the Slavnov--Taylor identity for the $V\bar c^+ c$ vertex. In the next section it will be used for proving its finiteness in all orders of the perturbation theory.

\subsection{All-loop finiteness of the triple gauge-ghost vertices}
\hspace*{\parindent}\label{Subsection_Finiteness}

To prove the all-loop finiteness of the triple gauge-ghost vertices in which all external lines correspond to the same non-Abelian subgroup $G_K$, we first prove the finiteness of the functions $H_K$ defined by Eq. (\ref{H_Definition}). Each of these functions is contributed by superdiagrams in which one external line corresponds to the source ${\cal J}$, and the other two correspond to the Faddeev--Popov ghost $c_K$. The vertex with the external source is given by the expression (\ref{JCC_Vertex}). If both ghost superfields in Eq. (\ref{JCC_Vertex}) produce propagators, then the supergraph contributing to the function $H_K$ contains

\begin{equation}\label{JCC_Vertex_With_Propagators}
\int d^4y\, d^2\theta_y\, {\cal J}_y^{A_K} \cdot \frac{\bar D_y^2  D_y^2}{4\partial^2} \delta^8_{y1} \cdot \frac{\bar D_y^2  D_y^2}{4\partial^2} \delta^8_{y2},
\end{equation}

\noindent
where two ghost propagators connect the superspace point $y$ to the superspace points $1$ and $2$. With the help of the identity

\begin{equation}\label{Superspace_Measure}
\int d^4x\, d^4\theta_x = - \frac{1}{2} \int d^4x\, d^2\theta_x\, \bar D_x^2
\end{equation}

\noindent
the expression (\ref{JCC_Vertex_With_Propagators}) can be presented in the form of the integral over the full superspace,

\begin{equation}
-2 \int d^4y\, d^4\theta_y\, {\cal J}_y^{A_K} \cdot \frac{D_y^2}{4\partial^2} \delta^8_{y1} \cdot \frac{\bar D_y^2  D_y^2}{4\partial^2} \delta^8_{y2}.
\end{equation}

\noindent
Evidently, if (only) one ghost superfield in the vertex (\ref{JCC_Vertex}) corresponds to an external line, then the vertex can also be presented as such an integral. (Two external ghosts in Eq. (\ref{JCC_Vertex}) can appear only in the tree contribution to the function $H_K$.) This implies that all vertices in the considered supergraph can be written as integrals over the full superspace. Therefore, it is possible to apply standard rules for calculating supergraphs. It is well known (see, e.g., \cite{West:1990tg}) that the resulting expression is also given by an integral over the full superspace. However, all external lines (which correspond to ${\cal J}_K$ and two $c_K$) are chiral. Therefore, due to Eq. (\ref{Superspace_Measure}) a nontrivial result can be obtained only if at least two right spinor covariant derivatives act on the external lines. If we take into account that the function $H_K$ is dimensionless, this implies that the dimension of the remaining loop integral is $m^{-2}$. In other words, the degree of divergence is equal to $(-2)$, so that this integral is finite in the ultraviolet region. Due to the renormalizability of the considered theory possible subdivergences are removed after the renormalization in the previous orders. Thus, the functions $H_K$ (for any $K$ such that $G_K\ne U(1)$) appear to UV finite.

Next, we involve the Slavnov--Taylor identity (\ref{STI_Identity_Final}) multiplied by the ghost renormalization constant $Z_{c_K}$. Using the finiteness of the renormalized Green function

\begin{equation}
(G_{c_K})_R(\alpha,\lambda,q^2/\mu^2) = \lim\limits_{\Lambda\to\infty} Z_{c_K}(\alpha,\lambda,\Lambda/\mu) G_{c_K}(\alpha_0,\lambda_0,q^2/\Lambda^2)
\end{equation}

\noindent
we see that

\begin{equation}
\lim\limits_{\Lambda\to\infty} \frac{d(Z_{c_K} G_{c_K})}{d\ln\Lambda} \bigg|_{\alpha,\lambda=\mbox{\scriptsize const}} = 0.
\end{equation}

\noindent
Therefore, differentiating Eq. (\ref{STI_Identity_Final}) multiplied by $Z_{c_K}$ with respect to $\ln\Lambda$ and keeping in mind the finiteness of the function $H_K$ we obtain

\begin{eqnarray}
&& 0 = \lim\limits_{\Lambda\to\infty} \frac{d}{d\ln\Lambda}\Big(Z_{c_K} G_{c_K}(q) {\cal S}_K(q,p) +  Z_{c_K} G_{c_K}(p) {\cal S}_K(p,q)\Big)\bigg|_{\alpha,\lambda=\mbox{\scriptsize const}}\nonumber\\
&&\qquad\qquad\qquad\qquad = \lim\limits_{\Lambda\to\infty}\Big(Z_{c_K} G_{c_K}(q) \frac{d}{d\ln\Lambda} {\cal S}_K(q,p) + Z_{c_K} G_{c_K}(p) \frac{d}{d\ln\Lambda} {\cal S}_K(p,q) \Big).\qquad
\end{eqnarray}

\noindent
In this equation we set $p=-q$ and take into account that $G_{c_K}(-q) = G_{c_K}(q)$ because this function depends only on $q^2/\Lambda^2$. The result can be written in the form

\begin{equation}
\lim\limits_{\Lambda\to\infty} \frac{d}{d\ln\Lambda} \Big({\cal S}_K(q,-q) + {\cal S}_K(-q,q) \Big) = 0.
\end{equation}

\noindent
The function ${\cal S}_K(-q,q)$ also depends only on $q^2/\Lambda^2$, so that from the above equation we conclude that it is finite in the ultraviolet region.

In our notation the renormalization constants for the gauge coupling constants, for the quantum gauge superfields, and for the Faddeev--Popov ghosts are defined by the equations

\begin{equation}
\frac{1}{\alpha_{0K}} = \frac{Z_{\alpha_K}}{\alpha_K}; \qquad V_K = Z_{V_K} Z_{\alpha_K}^{-1/2} (V_K)_R; \qquad \bar c_K c_K = Z_{c_K} Z_{\alpha_K}^{-1} (\bar c_K)_R (c_K)_R,
\end{equation}

\noindent
respectively. Taking into account that

\begin{eqnarray}
&&\hspace*{-3mm} V_K = e_{0K} V^{A_K} T^{A_K};\qquad\qquad\ \bar c_K = e_{0K} \bar c^{A_K} T^{A_K}; \qquad\quad\ \ \, c_K = e_{0K} c^{A_K} T^{A_K}; \vphantom{\Big(}\\
&&\hspace*{-3mm} \left(V_K\right)_R = e_K \left(V^{A_K}\right)_R T^{A_K};\quad\ \left(\bar c_K\right)_R = e_K \left(\bar c^{A_K}\right)_R T^{A_K}; \quad (c_K)_R = e_K \left(c^{A_K}\right)_R T^{A_K}\qquad\vphantom{\Big(}
\end{eqnarray}

\noindent
we see that the components of the quantum gauge superfields and of the Faddeev--Popov ghosts are renormalized as

\begin{equation}
V^{A_K} = Z_{V_K} \left(V^{A_K}\right)_R;\qquad \bar c^{A_K} c^{B_K} = Z_{c_K} \left(\bar c^{A_K}\right)_R \left(c^{B_K}\right)_R.
\end{equation}

\noindent
Using these equations one can obtain the renormalization constants for the triple gauge-ghost vertices from Eq. (\ref{S_Definition}). Certainly they are the same for all 4 types of the considered vertices and for all functions entering Eq. (\ref{S_Definition}). For the vertex corresponding to the subgroup $G_K$ the renormalization constant is equal to $Z_{\alpha_K}^{-1/2} Z_{c_K} Z_{V_K}$, for example,

\begin{equation}\label{S_Renormalization}
\left({\cal S}_K\right)_R(p,q) = \lim\limits_{\Lambda\to\infty} Z_{\alpha_K}^{-1/2} Z_{c_K} Z_{V_K} {\cal S}_K(p,q).
\end{equation}

\noindent
The renormalized function in the left hand side of this equation is UV finite by definition. Certainly, this is so for the particular case $p=-q$. From the other side, as we discussed above, the function ${\cal S}_K(-q,q)$ is not divergent in the ultraviolet region. Therefore, the corresponding renormalization constant should be finite,

\begin{equation}\label{ZZZ_Derivative}
\frac{d}{d\ln\Lambda} \Big(Z_{\alpha_K}^{-1/2} Z_{c_K} Z_{V_K}\Big) = 0.
\end{equation}

\noindent
This implies that it is possible (although not necessary) to choose a renormalization scheme in which the renormalization constants satisfy the relation

\begin{equation}\label{ZZZ_Relation}
Z_{\alpha_K}^{-1/2} Z_{c_K} Z_{V_K} = 1.
\end{equation}

\noindent
Evidently, this equation holds in the HD+MSL scheme, because in this case various $Z$-s include only powers of $\ln\Lambda/\mu$, while all finite constants vanish.

\subsection{A new form of the NSVZ $\beta$-function for theories with multiple gauge couplings}
\hspace*{\parindent}\label{Subsection_New_NSVZ}

Let us derive a new form of the NSVZ equation for theories with multiple gauge couplings making the transformations analogous to the ones proposed in Ref. \cite{Stepanyantz:2016gtk}. However, here we will use RGFs defined in terms of the renormalized coupling assuming that Eq. (\ref{ZZZ_Relation}) is satisfied. Certainly, for RGFs defined in terms of the bare couplings a similar result is also valid. (In this case it is sufficient to use Eq. (\ref{ZZZ_Derivative}), while Eq. (\ref{ZZZ_Relation}) is not needed.)

First, we equivalently rewrite the NSVZ equation (\ref{NSVZ_Multicharge}) in the form

\begin{equation}\label{NSVZ_Equaivalent}
\frac{\beta_K(\alpha,\lambda)}{\alpha_K^2} = - \frac{1}{2\pi} \Big[\, 3 C_2(G_K) - \sum\limits_{\mbox{\scriptsize a}} \bm{T}_{\mbox{\scriptsize a}K}\Big(1-\gamma_{\mbox{\scriptsize a}}(\alpha,\lambda)\Big) \Big] + \frac{C_2(G_K)}{2\pi}\cdot \frac{\beta_K(\alpha,\lambda)}{\alpha_K}.
\end{equation}

\noindent
It is convenient to express the function $\beta_K(\alpha,\lambda)$ in the right hand of this equation in terms of the renormalization constant for the gauge coupling $\alpha_K$,

\begin{equation}
\beta_K(\alpha,\lambda) = \frac{d\alpha_K}{d\ln\mu}\bigg|_{\alpha_0,\lambda_0=\mbox{\scriptsize const}} = \alpha_K \frac{d\ln Z_{\alpha_K}}{d\ln\mu}\bigg|_{\alpha_0,\lambda_0=\mbox{\scriptsize const}}.
\end{equation}

\noindent
This renormalization constant can in turn be related to the renormalization constants of the quantum gauge superfield and of the Faddeev--Popov ghosts corresponding to the subgroup $G_K$ with the help of equation (\ref{ZZZ_Relation}). As a result, the $\beta$-function for a certain gauge coupling can be expressed in terms of the anomalous dimensions for the corresponding quantum gauge superfield and Faddeev--Popov ghosts,

\begin{equation}\label{Beta_In_Term_Of_Gammas}
\beta_K(\alpha,\lambda) = 2\alpha_K \frac{d}{d\ln\mu}\ln\left(Z_{V_K} Z_{c_K}\right) = 2\alpha_K \Big(\gamma_{V_K}(\alpha,\lambda) + \gamma_{c_K}(\alpha,\lambda)\Big),
\end{equation}

\noindent
where

\begin{equation}
\gamma_{V_K}(\alpha,\lambda) = \frac{d\ln Z_{V_K}}{d\ln\mu}\bigg|_{\alpha_0,\lambda_0=\mbox{\scriptsize const}};\qquad \gamma_{c_K}(\alpha,\lambda) = \frac{d\ln Z_{c_K}}{d\ln\mu}\bigg|_{\alpha_0,\lambda_0=\mbox{\scriptsize const}}.
\end{equation}

\noindent
Substituting the expression (\ref{Beta_In_Term_Of_Gammas}) for the function $\beta_K$ into the right hand side of Eq. (\ref{NSVZ_Equaivalent}) we obtain the equivalent form the NSVZ relation

\begin{equation}\label{NSVZ_Multicharge_New}
\frac{\beta_K(\alpha,\lambda)}{\alpha_K^2} = - \frac{1}{2\pi} \Big[\, C_2(G_K) \Big(3 - 2 \gamma_{V_K}(\alpha,\lambda) - 2\gamma_{c_K}(\alpha,\lambda)\Big)  - \sum\limits_{\mbox{\scriptsize a}} \bm{T}_{\mbox{\scriptsize a}K}\Big(1-\gamma_{\mbox{\scriptsize a}}(\alpha,\lambda)\Big) \Big],
\end{equation}

\noindent
which generalizes a similar equation constructed in \cite{Stepanyantz:2016gtk} for theories with a single gauge coupling constant.

Unlike the original NSVZ relation (\ref{NSVZ_Multicharge}), this form of the NSVZ equation has a simple graphical interpretation. If we consider a vacuum supergraph (which does not contain external lines), then the corresponding contribution to the function $\beta_K$ is produced by the sum of all superdiagrams which are obtained by attaching two external lines of the background gauge superfield $\bm{V}_K$. From the other side, cutting internal lines we produce superdiagrams contributing to anomalous dimensions of various quantum superfields. Eq. (\ref{NSVZ_Multicharge_New}) relates these contributions to the $\beta$-function.\footnote{For the quantum gauge superfield this is correct only if the corresponding contribution to the two-point Green function is transversal. Certainly, due to the Slavnov--Taylor identities this is always so for the sum of all supergraphs.} However, it is worth to note that for the quantum gauge superfields only cuts of the $V_K$-propagators contribute to the function $\beta_K$. Possibly, this can be explained with the help of the method for constructing integrals of double total derivatives proposed in \cite{Stepanyantz:2019ihw}. We also hope that the generalization of the results obtained in \cite{Stepanyantz:2020uke} allows deriving Eq. (\ref{NSVZ_Multicharge_New}) in a way analogous to the case of theories with a single gauge coupling constant.

\section*{Conclusion}
\hspace*{\parindent}

In this paper we investigated the NSVZ equations for ${\cal N}=1$ supersymmetric models with multiple gauge couplings. In these theories the gauge group is a direct product of simple subgroups and $U(1)$ factors. Certainly, there are a lot of various phenomenologically interesting examples of such theories, for instance, MSSM or flipped $SU(5)$ GUT considered in this paper in detail. In theories with multiple gauge couplings a number of $\beta$-functions is equal to a number of (simple and $U(1)$) factors in the gauge group. We assume that in all loops each of these $\beta$-functions satisfies the NSVZ equations (\ref{NSVZ_Multicharge}) and (\ref{NSVZ_Multicharge_New}) in a certain subtraction scheme. A strong evidence in favor of this is that in MSSM and flipped $SU(5)$ GUT predictions of the NSVZ equations for the two-loop $\beta$-functions exactly coincide with the corresponding (well-known) results obtained by direct calculations made in $\overline{\mbox{DR}}$-scheme. Certainly, this coincidence occurs because the two-loop gauge $\beta$-functions are scheme independent.\footnote{Note that the Yukawa $\beta$-functions are scheme dependent starting from the two-loop approximation.} From the NSVZ equations for theories with multiple gauge couplings it is also possible to derive the exact NSVZ-like equation for the Adler $D$-function in ${\cal N}=1$ SQCD, which was first obtained in \cite{Shifman:2014cya,Shifman:2015doa}.

It is highly probable that the NSVZ scheme for theories with multiple gauge couplings is given by the HD+MSL prescription as for theories with a simple gauge group \cite{Stepanyantz:2020uke}. HD means that a theory should be regularized by Higher covariant Derivatives. This regularization in particular includes insertion of the Pauli--Villars determinants for removing one-loop divergences. Note that the ratios of the Pauli--Villars masses to the dimensionful regularization parameter ($\Lambda$) should be independent of couplings. MSL indicates that divergences are removed with the help of Minimal Subtractions of Logarithms, when only powers of $\ln\Lambda/\mu$ are included into renormalization constants. Therefore, for investigating the NSVZ equations it is highly desirable to construct a version of the higher covariant derivative regularization applicable for theories multiple gauge couplings. This was also done in this paper for the general case and was illustrated by the examples of MSSM and flipped $SU(5)$ GUT. Also the higher covariant derivative regularization is a very important ingredient needed for the all-loop perturbative derivation of the NSVZ $\beta$-function. According to \cite{Stepanyantz:2016gtk,Stepanyantz:2019ihw,Stepanyantz:2019lfm,Stepanyantz:2020uke}, summing the perturbation series we do not obtain the original form of the NSVZ equation. Instead of it we obtain the relation between the $L$-loop $\beta$-function and the $(L-1)$-loop anomalous dimensions of the quantum gauge superfield, of the Faddeev--Popov ghosts, and of the matter superfields. The original NSVZ equation can be derived from it with the help of the non-renormalization theorem for the triple gauge-ghost vertices. In this paper we generalized this theorem to the case of theories with multiple gauge couplings and obtained from it the new form of the NSVZ equations for theories with multiple gauge couplings, which is given by Eq. (\ref{NSVZ_Multicharge_New}). We hope to finalize the all-loop perturbative derivation of these equations in the forthcoming publications.

\section*{Acknowledgements}
\hspace*{\parindent}

K.S. is very grateful to A.L.Kataev, R.Shrock, and P.West for indicating some important references.

This work was supported by the Foundation for the Advancement of Theoretical Physics and Mathematics `BASIS', grants No. 19-1-1-45-1 (K.S.) and 19-1-1-45-2 (N.T.).

\appendix

\section{NSVZ equations in the order $O(\alpha,\lambda^2)$}
\hspace*{\parindent}\label{Appendix_Two-Loop}

In this appendix we present expressions for the (scheme-indepedent) one-loop anomalous dimensions for all chiral matter superfields in MSSM and flipped $SU(5)$ GUT. We substitute them into the NSVZ equations and obtain  two-loop expressions for the $\beta$-functions. As we discussed in Sect. \ref{Subsection_Scheme_Dependence}, they do not depend on a renormalization prescription, so that it is possible to compare them with the known results obtained in the $\overline{\mbox{DR}}$-scheme. Below we demonstrate the agreement of the results derived by these two methods. (For MSSM this was first done in \cite{Ghilencea:1997mu}.) This agreement can be considered as a strong evidence that NSVZ equations discussed in this paper are really valid for theories with multiple gauge couplings.

\subsection{MSSM}
\hspace*{\parindent}\label{Subappendix_Two-Loop_MSSM}

Let us construct two-loop contributions to the $\beta$-functions of MSSM starting from the exact $\beta$-functions (\ref{MSSM_NSVZ_Beta3}) -- (\ref{MSSM_NSVZ_Beta1}). For this purpose we need expressions for the one-loop anomalous dimensions of all chiral matter superfields. For a theory with a simple gauge group the one-loop anomalous dimension is given by the expression

\begin{equation}\label{Gamma_Simple}
\gamma_i{}^j(\alpha,\lambda) = - \frac{\alpha}{\pi} C(R)_i{}^j + \frac{1}{4\pi^2} \lambda^*_{imn} \lambda^{jmn} + O(\alpha^2,\alpha\lambda^2,\lambda^4).
\end{equation}

\noindent
To generalize this result to the case of theories with multiple gauge couplings, we first discuss the structure of the matrix $\lambda^*_{imn} \lambda^{jmn}$. It is evidently Hermitian and invariant under the transformations of the gauge group. If matter superfields lie in an irreducible representation of a simple gauge group, then this matrix is proportional to $\delta_i^j$. However, in the general case it is not so. Nevertheless, making a proper unitary rotation in the generation space it is always possible to diagonalize the matrix $\lambda^*_{imn} \lambda^{jmn}$. We will always assume that this is done. When this matrix becomes diagonal, the superfields $\phi_{\mbox{\scriptsize a}}$ are its eigenvectors corresponding to the certain eigenvalues, which we will denote by $(\lambda^*\lambda)_{\mbox{\scriptsize a}}$. Then we see that for theories with multiple gauge couplings the one-loop expression for the anomalous dimension of the superfield $\phi_{\mbox{\scriptsize a}}$ generalizing Eq. (\ref{Gamma_Simple}) can be written in the form

\begin{equation}\label{One-Loop_Gamma_Multicharge}
\gamma_{\mbox{\scriptsize a}}(\alpha,\lambda) = - \sum\limits_K \frac{\alpha_K}{\pi} C(R_{\mbox{\scriptsize a} K}) + \frac{1}{4\pi^2} (\lambda^*\lambda)_{\mbox{\scriptsize a}} + O(\alpha^2,\alpha\lambda^2,\lambda^4),
\end{equation}

\noindent
where the factor $C(R_{\mbox{\scriptsize a} K})$ is defined by Eq. (\ref{C(R)_Definition}). These factors can be calculated as follows. If $R$ is an irreducible representation of a simple group $G$, then

\begin{equation}\label{C(R)_Calculation}
C(R) = \frac{1}{\mbox{dim}\,R}\, \mbox{tr} (T^A T^A) = \frac{ T(R)\,r}{\mbox{dim}\,R},
\end{equation}

\noindent
where $r$ is a dimension of the group $G$ and $\mbox{dim}\,R$ is a dimension of the representation $R$. In particular, for the (anti)fundamental representation of the group $SU(N)$

\begin{equation}
C\Big(\mbox{fund.}\, SU(N)\Big) = \frac{N^2-1}{2N}.
\end{equation}

\noindent
For the $U(1)$ subgroups $C(R_{\mbox{\scriptsize a} K})$ should be replaced by $q_{\mbox{\scriptsize a}}^2$, where $q_{\mbox{\scriptsize a}}$ is a charge of the superfield $\phi_{\mbox{\scriptsize a}}$ with respect to the considered $U(1)$ subgroup.

Because Eq. (\ref{One-Loop_Gamma_Multicharge}) contains the Yukawa couplings, for calculating one-loop anomalous dimensions we also need the explicit expression for a part of the MSSM action containing the superpotential. It is written in the form

\begin{equation}
\Delta S = \frac{1}{2} \int d^4x\, d^2\theta\, W(\phi_{\mbox{\scriptsize a}}) + \mbox{c.c.},
\end{equation}

\noindent
where the superpotential is given by the expression

\begin{eqnarray}
&&\hspace*{-7mm} W(\phi_{\mbox{\scriptsize a}}) = \left(Y_U\right)_{IJ}
\left(\widetilde U\ \widetilde D \right)^{a}_I
\left(
\begin{array}{cc}
0 & 1\\
-1 & 0
\end{array}
\right)
\left(
\begin{array}{c}
H_{u1}\\ H_{u2}
\end{array}
\right) U_{aJ}
+ \left(Y_D\right)_{IJ}
\left(\widetilde U\ \widetilde D \right)^{a}_I
\left(
\begin{array}{cc}
0 & 1\\
-1 & 0
\end{array}
\right)
\left(
\begin{array}{c}
H_{d1}\\ H_{d2}
\end{array}
\right)
\nonumber\\
&&\hspace*{-7mm} \times D_{aJ} + \left(Y_E\right)_{IJ} \left(\widetilde N\ \widetilde E \right)_{I}
\left(
\begin{array}{cc}
0 & 1\\
-1 & 0
\end{array}
\right)
\left(
\begin{array}{c}
H_{d1}\\ H_{d2}
\end{array}
\right) E_J
+ \bm{\mu} \left(H_{u1}\ H_{u2} \right)
\left(
\begin{array}{cc}
0 & 1\\
-1 & 0
\end{array}
\right)
\left(
\begin{array}{c}
H_{d1}\\ H_{d2}
\end{array}
\right).
\end{eqnarray}

\noindent
In this expression the indices $I,J =1,2,3$ numerate generations of elementary particles. The Yukawa coupling $\lambda^{ijk}$ can be written in terms of the $3\times 3$ Yukawa matrices $\left(Y_U\right)_{IJ}$, $\left(Y_D\right)_{IJ}$, and $\left(Y_E\right)_{IJ}$. The parameter $\bm{\mu}$ (which should be distinguished from the normalization point $\mu$) has the dimension of mass and is not essential for calculating the considered renormalization group functions.

Now, it is possible to obtain all expressions for the anomalous dimensions of the chiral matter superfields starting from Eq. (\ref{One-Loop_Gamma_Multicharge}). Certainly, constructing them we need to take into account the factor $5/3$ present in the definition of the constant $\alpha_1$, see Eq. (\ref{Couplings_Definitions}). The result can be written in form (see, e.g., \cite{Jack:2004ch})

\begin{eqnarray}\label{MSSM_Anomalous_Dimantions_Q}
&& \gamma_Q(\alpha,Y) = - \frac{\alpha_1}{60\pi} - \frac{3\alpha_2}{4\pi} - \frac{4\alpha_3}{3\pi} + \frac{1}{8\pi^2}\Big(Y_U Y_U^+ + Y_D Y_D^+\Big) + O(\alpha^2,\alpha Y^2, Y^4);\qquad\\
&& \gamma_U(\alpha,Y) = - \frac{4 \alpha_1}{15 \pi} - \frac{4\alpha_3}{3\pi} + \frac{1}{4\pi^2}\, Y_U^+ Y_U + O(\alpha^2,\alpha Y^2, Y^4);\\
&& \gamma_D(\alpha,Y) = - \frac{\alpha_1}{15\pi} - \frac{4\alpha_3}{3\pi} + \frac{1}{4\pi^2}\, Y_D^+ Y_D + O(\alpha^2,\alpha Y^2, Y^4);\\
&& \gamma_L(\alpha,Y) = - \frac{3\alpha_1}{20\pi} - \frac{3\alpha_2}{4\pi} + \frac{1}{8\pi^2}\, Y_E Y_E^+  + O(\alpha^2,\alpha Y^2, Y^4);\\
&& \gamma_E(\alpha,Y) = - \frac{3\alpha_1}{5\pi} + \frac{1}{4\pi^2}\, Y_E^+ Y_E + O(\alpha^2,\alpha Y^2, Y^4);\\
&& \gamma_{H_u}(\alpha,Y) = - \frac{3\alpha_1}{20\pi} - \frac{3\alpha_2}{4\pi} + \frac{3}{8\pi^2}\, \mbox{tr}\Big(Y_U^+ Y_U\Big) + O(\alpha^2,\alpha Y^2, Y^4);\\
\label{MSSM_Anomalous_Dimantions_Hd}
&& \gamma_{H_d}(\alpha,Y) = - \frac{3\alpha_1}{20\pi} - \frac{3\alpha_2}{4\pi} + \frac{1}{8\pi^2}\, \mbox{tr}\Big(3\, Y_D^+ Y_D + Y_E^+ Y_E\Big) + O(\alpha^2,\alpha Y^2, Y^4).
\end{eqnarray}

\noindent
Note that in our conventions the matrices $Y_U^+ Y_U$, $Y_D^+ Y_D$, $Y_E^+ Y_E$ should be made diagonal by specially constructed unitary rotations of chiral superfields $U$, $D$, and $E$ which act on the indices numerating generations. Similarly, the matrices $Y_U Y_U^+ + Y_D Y_D^+$ and $Y_E Y_E^+$ should be made diagonal by proper unitary rotations of the superfields $Q$ and $L$, respectively.\footnote{Usually such rotations are chosen in such a way that certain Yukawa matrices will be diagonal, real, and positive. The convention adopted in this paper certainly differs from it.} In Eqs. (\ref{MSSM_Anomalous_Dimantions_Q}) --- (\ref{MSSM_Anomalous_Dimantions_Hd}) we certainly assume that for calculating anomalous dimensions of the quark and lepton superfields it is necessary to take a diagonal element of a proper product of the Yukawa matrices which corresponds to the considered generation.

Substituting Eqs. (\ref{MSSM_Anomalous_Dimantions_Q}) --- (\ref{MSSM_Anomalous_Dimantions_Hd}) into the NSVZ equations (\ref{MSSM_NSVZ_Beta3}) --- (\ref{MSSM_NSVZ_Beta1}) after some simple algebra we obtain the two-loop MSSM $\beta$-functions,

\begin{eqnarray}
&& \frac{\beta_3(\alpha,Y)}{\alpha_3^2} = - \frac{1}{2\pi} \bigg[3 -\frac{11\alpha_1}{20\pi} -\frac{9\alpha_2}{4\pi} -\frac{7\alpha_3}{2\pi}
\nonumber\\
&&\qquad\qquad\qquad\qquad\qquad\qquad + \frac{1}{8\pi^2}\, \mbox{tr}\Big(2\, Y_U^+ Y_U + 2\, Y_D^+ Y_D\Big) \bigg] + O(\alpha^2,\alpha Y^2, Y^4);\qquad\\
&& \frac{\beta_2(\alpha,Y)}{\alpha_2^2} = - \frac{1}{2\pi} \bigg[-1  -\frac{9\alpha_1}{20\pi}- \frac{25\alpha_2}{4\pi} -\frac{6\alpha_3}{\pi} \nonumber\\
&&\qquad\qquad\qquad\qquad\, + \frac{1}{8\pi^2}\, \mbox{tr}\Big(3\, Y_U^+ Y_U + 3\, Y_D^+ Y_D + Y_E^+ Y_E \Big) \bigg]+ O(\alpha^2,\alpha Y^2, Y^4);\\
&& \frac{\beta_1(\alpha,Y)}{\alpha_1^2} = - \frac{1}{2\pi}\bigg[-\frac{33}{5} -\frac{199\alpha_1}{100\pi} -\frac{27\alpha_2}{20\pi} -\frac{22\alpha_3}{5\pi}
\nonumber\\
&&\qquad\qquad\quad\ \ + \frac{1}{8\pi^2}\cdot \frac{1}{5}\, \mbox{tr}\Big(13\, Y_U^+ Y_U + 7\, Y_D^+ Y_D + 9\, Y_E^+ Y_E \Big) \bigg]+ O(\alpha^2,\alpha Y^2, Y^4).
\end{eqnarray}

\noindent
They exactly coincide with the known expressions (which can be found in, e.g., \cite{Martin:1993zk}).

\subsection{The flipped $SU(5)$ model}
\hspace*{\parindent}\label{Subappendix_Two-Loop_Flipped}

Now let us calculate two-loop $\beta$-functions for the flipped $SU(5)$ model by substituting one-loop anomalous dimensions of various chiral superfields into the NSVZ equations. For completeness, here we consider the model containing $N_G$ generations of the matter superfields in the representation $\overline{10}(1)+5(-3)+1(5)$, $N_H$ Higgs superfields in the representation $10(-1)+\overline{10}(1)$, $N_h$ Higgs superfields in the representation $5(2) + \overline{5}(-2)$, and $N_G+1$ singlets $\phi$. The theory considered in Sect. \ref{Subsection_Flipped_NSVZ} corresponds to

\begin{equation}
N_G =3;\qquad N_H =1;\qquad N_h = 1.
\end{equation}

\noindent
Below indices of the chiral superfields $\overline{10}^{\,ij}$, $5_i$, and $E$ which numerate the generations will be denoted by the letters $I,J = 1,\ldots,N_G$. Similarly, the Higgs superfields $h_i$ and $\widetilde h^i$ will be numerated by the letters $K,L=1,\ldots,N_h$. For the Higgs superfields $H_{ij}$ and $\widetilde H^{ij}$ we will use the letters $M,N=1,\ldots,N_H$. The singlets will be numerated by the letters $R,S,T=1,\ldots,N_G+1$. The superpotential of the model has the form \cite{Antoniadis:1987dx}

\begin{eqnarray}
&&\hspace*{-5mm} W(\phi_{\mbox{\scriptsize a}}) = \left(\lambda_1\right)^{IJK} \varepsilon_{ijklm} (\overline{10}^{\,ij})_I (\overline{10}^{\,kl})_J  (\widetilde h^m)_K + \left(\lambda_2\right)^{IJK} (\overline{10}^{\,ij})_I (5_{i})_J (h_j)_K + \left(\lambda_3\right)^{IJK} (5_{i})_I \vphantom{\Big(}\nonumber\\
&&\hspace*{-5mm} \times\, E_J (\widetilde h^i)_K + \left(\lambda_4\right)^{MNK}\, \varepsilon_{ijklm} (\widetilde H^{ij})_M (\widetilde H^{kl})_N (\widetilde h^m)_K + \left(\lambda_5\right)^{MNK}\, \varepsilon^{ijklm} (H_{ij})_M (H_{kl})_N (h_{m})_K
\vphantom{\Big(}\nonumber\\
&&\hspace*{-5mm} + \left(\lambda_6\right)^{IMR} (\overline{10}^{\,ij})_I (H_{ij})_M \phi_R + \left(\lambda_7\right)^{KLR} (h_i)_K (\widetilde h^i)_L \phi_R + \left(\lambda_8\right)^{RST} \phi_R\phi_S\phi_T.  \vphantom{\Big(}
\end{eqnarray}

\noindent
Note that in our notation the superfields in the representations $10$ and $\overline{10}$ are normalized in such a way that their kinetic terms contain the coefficient $1/4$. For instance, in the classical action the term quadratic in the superfield $H_{ij}$ (without the gauge superfields) has the form

\begin{equation}
\frac{1}{4} \int d^4x\,d^4\theta\, (H^*)^{ij} H_{ij}.
\end{equation}

\noindent
To avoid double summations, one can rescale all superfields in the representations $10$ and $\overline{10}$ in $1/\sqrt{2}$ times, e.g., $H_{ij} \to H_{ij}/\sqrt{2}$, etc.

If $N_G$, $N_H$, and $N_h$ are not fixed, then from Eq. (\ref{NSVZ_Multicharge}) and Table \ref{Table_Flipped_T} (and also using Eq. (\ref{Flipped_Alpha_1})) we obtain the exact $\beta$-functions

\begin{eqnarray}\label{Fliiped_Total_Beta5}
&&\hspace*{-5mm} \frac{\beta_5(\alpha,\lambda)}{\alpha_5^2} =  - \frac{1}{2\pi(1 - 5\alpha_5/2\pi)} \bigg[\, 15 - 2 N_G - 3 N_H - N_h  +  \sum\limits_{I=1}^{N_G}\Big(\frac{3}{2} \gamma_{\overline{10}_I}(\alpha,\lambda) +\frac{1}{2} \gamma_{5_I}(\alpha,\lambda)\Big) \nonumber\\
&&\hspace*{-5mm} + \frac{3}{2} \sum\limits_{M=1}^{N_H} \Big(\gamma_{H_M}(\alpha,\lambda) + \gamma_{\widetilde{H}_M}(\alpha,\lambda)\Big)
+ \frac{1}{2} \sum\limits_{K=1}^{N_h} \Big(\gamma_{h_K}(\alpha,\lambda) + \gamma_{\widetilde{h}_K}(\alpha,\lambda)\Big)\bigg]; \\
\label{Fliiped_Total_Beta1}
&&\hspace*{-5mm} \frac{\beta_1(\alpha,\lambda)}{\alpha_1^2} = \frac{1}{2\pi}\bigg[\, 2 N_G + \frac{1}{2} N_H + N_h - \frac{1}{8} \sum\limits_{I=1}^{N_G}\Big(2 \gamma_{\overline{10}_I}(\alpha,\lambda)  + 9 \gamma_{5_I}(\alpha,\lambda) +5 \gamma_{E_I}(\alpha,\lambda)\Big)  \nonumber\\
&&\hspace*{-5mm} - \frac{1}{4} \sum\limits_{M=1}^{N_H} \Big(\gamma_{H_M}(\alpha,\lambda) + \gamma_{\widetilde{H}_M}(\alpha,\lambda)\Big) - \frac{1}{2} \sum\limits_{K=1}^{N_h} \Big(\gamma_{h_K}(\alpha,\lambda) + \gamma_{\widetilde{h}_K}(\alpha,\lambda)\Big)\,\bigg].
\end{eqnarray}

To construct the anomalous dimensions entering these equations, we use Eq. (\ref{One-Loop_Gamma_Multicharge}). With the help of Eq. (\ref{C(R)_Calculation}) we see that the relevant coefficients $C(R_{\mbox{\scriptsize a} K})$ present in the part of this equation containing the gauge coupling constants have the form

\begin{equation}
C(5) = C(\overline{5}) = \frac{1}{2}\cdot \frac{24}{5} = \frac{12}{5}; \qquad C(10) = C(\overline{10}) = \frac{3}{2}\cdot \frac{24}{10} = \frac{18}{5}.
\end{equation}

\noindent
Using these values after some calculations we obtain the anomalous dimensions for all chiral superfields of the model. In the expressions for them listed below, the summation is not performed over the indices indicated in bold letters,

\begin{eqnarray}
&&\hspace*{-9mm} \gamma_{\overline{10}_I}(\alpha,\lambda) = - \frac{\alpha_1}{40\pi} - \frac{18\alpha_5}{5\pi} + \frac{6}{\pi^2} \left(\lambda_1^*\right)_{\bm{I}JK} \left(\lambda_1\right)^{\bm{I}JK} + \frac{1}{8\pi^2} \left(\lambda_2^*\right)_{\bm{I}JK} \left(\lambda_2\right)^{\bm{I}JK}\nonumber\\
&&\qquad\qquad\qquad\qquad\qquad\qquad\qquad\quad\, + \frac{1}{8\pi^2} \left(\lambda_6^*\right)_{\bm{I}MR} \left(\lambda_6\right)^{\bm{I}MR} + O(\alpha^2,\alpha\lambda^2,\lambda^4);\\
&&\hspace*{-9mm} \gamma_{5_I}(\alpha,\lambda) = - \frac{9\alpha_1}{40\pi} - \frac{12\alpha_5}{5\pi} + \frac{1}{4\pi^2} \left(\lambda_2^*\right)_{J\bm{I}K} \left(\lambda_2\right)^{J\bm{I}K}\nonumber\\
&&\hspace*{-9mm} \qquad\qquad\qquad\qquad\qquad\qquad\qquad\qquad\quad\ \, + \frac{1}{8\pi^2} \left(\lambda_3^*\right)_{\bm{I}JK} \left(\lambda_3\right)^{\bm{I}JK} + O(\alpha^2,\alpha\lambda^2,\lambda^4);\\
&&\hspace*{-9mm} \gamma_{E_I}(\alpha,\lambda) = - \frac{5\alpha_1}{8\pi}  + \frac{5}{8\pi^2} \left(\lambda_3^*\right)_{J\bm{I}K} \left(\lambda_3\right)^{J\bm{I}K} + O(\alpha^2,\alpha\lambda^2,\lambda^4);\\
&&\hspace*{-9mm} \gamma_{H_M}(\alpha,\lambda) = - \frac{\alpha_1}{40\pi} - \frac{18\alpha_5}{5\pi} + \frac{6}{\pi^2} \left(\lambda_5^*\right)_{\bm{M}NK} \left(\lambda_5\right)^{\bm{M}NK}\nonumber\\
&&\hspace*{-9mm} \qquad\qquad\qquad\qquad\qquad\qquad\qquad\qquad\quad + \frac{1}{8\pi^2} \left(\lambda_6^*\right)_{I\bm{M}R} \left(\lambda_6\right)^{I\bm{M}R} + O(\alpha^2,\alpha\lambda^2,\lambda^4);\\
&&\hspace*{-9mm} \gamma_{\widetilde H_M}(\alpha,\lambda) = - \frac{\alpha_1}{40\pi} - \frac{18\alpha_5}{5\pi} + \frac{6}{\pi^2} \left(\lambda_4^*\right)_{\bm{M}NK} \left(\lambda_4\right)^{\bm{M}NK}
+ O(\alpha^2,\alpha\lambda^2,\lambda^4);\\
&&\hspace*{-9mm} \gamma_{h_K}(\alpha,\lambda) = - \frac{\alpha_1}{10\pi} - \frac{12\alpha_5}{5\pi}  + \frac{1}{4\pi^2} \left(\lambda_2^*\right)_{IJ\bm{K}} \left(\lambda_2\right)^{IJ\bm{K}}\nonumber\\
&&\hspace*{-9mm} \qquad\qquad\qquad + \frac{6}{\pi^2} \left(\lambda_5^*\right)_{MN\bm{K}} \left(\lambda_5\right)^{MN\bm{K}} + \frac{1}{8\pi^2} \left(\lambda_7^*\right)_{\bm{K}LR} \left(\lambda_7\right)^{\bm{K}LR} + O(\alpha^2,\alpha\lambda^2,\lambda^4);\\
&&\hspace*{-9mm} \gamma_{\widetilde h_K}(\alpha,\lambda) = - \frac{\alpha_1}{10\pi} - \frac{12\alpha_5}{5\pi} + \frac{6}{\pi^2} \left(\lambda_1^*\right)_{IJ\bm{K}} \left(\lambda_1\right)^{IJ\bm{K}}
+ \frac{1}{8\pi^2} \left(\lambda_3^*\right)_{IJ\bm{K}} \left(\lambda_3\right)^{IJ\bm{K}} \nonumber\\
&&\hspace*{-9mm}\qquad\qquad\qquad + \frac{6}{\pi^2} \left(\lambda_4^*\right)_{MN\bm{K}} \left(\lambda_4\right)^{MN\bm{K}} + \frac{1}{8\pi^2} \left(\lambda_7^*\right)_{L\bm{K}R} \left(\lambda_7\right)^{L\bm{K}R}
+ O(\alpha^2,\alpha\lambda^2,\lambda^4);\\
&&\hspace*{-9mm} \gamma_{\phi_R}(\alpha,\lambda) = \frac{5}{4\pi^2} \left(\lambda_6^*\right)_{IM\bm{R}} \left(\lambda_6\right)^{IM\bm{R}} + \frac{5}{8\pi^2} \left(\lambda_7^*\right)_{KL\bm{R}} \left(\lambda_7\right)^{KL\bm{R}} \nonumber\\
&&\hspace*{-9mm} \qquad\qquad\qquad\qquad\qquad\qquad\qquad\qquad\quad\ \, + \frac{9}{4\pi^2} \left(\lambda_8^*\right)_{\bm{R}ST} \left(\lambda_8\right)^{\bm{R}ST} + O(\alpha^2,\alpha\lambda^2,\lambda^4).
\end{eqnarray}

\noindent
Substituting these anomalous dimensions into the NSVZ equations (\ref{Fliiped_Total_Beta5}) and (\ref{Fliiped_Total_Beta1}) we obtain the two-loop expressions for the $\beta$-functions of the model under consideration,

\begin{eqnarray}
&&\hspace*{-7mm} \frac{\beta_5(\alpha,\lambda)}{\alpha_5^2} =  - \frac{1}{2\pi} \bigg[\, 15 - 2 N_G - 3 N_H - N_h +\frac{75\alpha_5}{2\pi} - N_G \Big(\frac{3\alpha_1}{20\pi} + \frac{58\alpha_5}{5\pi}\Big)
- N_H \Big(\frac{3\alpha_1}{40\pi} + \frac{183\alpha_5}{10\pi}\Big) \nonumber\\
&&\hspace*{-7mm}  - N_h \Big(\frac{\alpha_1}{10\pi} + \frac{49\alpha_5}{10\pi}\Big)
+ \frac{12}{\pi^2} \left(\lambda_1^*\right)_{IJK} \left(\lambda_1\right)^{IJK} + \frac{7}{16\pi^2} \left(\lambda_2^*\right)_{IJK} \left(\lambda_2\right)^{IJK}
+ \frac{1}{8\pi^2} \left(\lambda_3^*\right)_{IJK} \left(\lambda_3\right)^{IJK} \nonumber\\
&&\hspace*{-7mm}  + \frac{12}{\pi^2} \left(\lambda_4^*\right)_{MNK} \left(\lambda_4\right)^{MNK}
+ \frac{12}{\pi^2} \left(\lambda_5^*\right)_{MNK} \left(\lambda_5\right)^{MNK}  + \frac{3}{8\pi^2} \left(\lambda_6^*\right)_{IMR} \left(\lambda_6\right)^{IMR}
+ \frac{1}{8\pi^2} \left(\lambda_7^*\right)_{KLR}  \nonumber\\
&&\hspace*{-7mm} \times \left(\lambda_7\right)^{KLR}\bigg] + O(\alpha^2,\alpha\lambda^2,\lambda^4); \\
&& \vphantom{1}\nonumber\\
&&\hspace*{-7mm} \frac{\beta_1(\alpha,\lambda)}{\alpha_1^2} = -\frac{1}{2\pi}\bigg[ - 2 N_G - \frac{1}{2} N_H - N_h - N_G\Big(\frac{13\alpha_1}{20\pi} + \frac{18\alpha_5}{5\pi}\Big)
- N_H \Big(\frac{\alpha_1}{80\pi} + \frac{9\alpha_5}{5\pi}\Big) - N_h \Big(\frac{\alpha_1}{10\pi} \nonumber\\
&&\hspace*{-7mm}  + \frac{12\alpha_5}{5\pi}\Big) + \frac{9}{2\pi^2} \left(\lambda_1^*\right)_{IJK} \left(\lambda_1\right)^{IJK} + \frac{7}{16\pi^2} \left(\lambda_2^*\right)_{IJK} \left(\lambda_2\right)^{IJK} + \frac{19}{32\pi^2} \left(\lambda_3^*\right)_{IJK} \left(\lambda_3\right)^{IJK} + \frac{9}{2\pi^2} \nonumber\\
&&\hspace*{-7mm} \times \left(\lambda_4^*\right)_{MNK} \left(\lambda_4\right)^{MNK}
+ \frac{9}{2\pi^2} \left(\lambda_5^*\right)_{MNK} \left(\lambda_5\right)^{MNK} + \frac{1}{16\pi^2} \left(\lambda_6^*\right)_{IMR} \left(\lambda_6\right)^{IMR}
+ \frac{1}{8\pi^2} \left(\lambda_7^*\right)_{KLR}  \nonumber\\
&&\hspace*{-7mm} \times \left(\lambda_7\right)^{KLR}\bigg] + O(\alpha^2,\alpha\lambda^2,\lambda^4).
\end{eqnarray}

\noindent
The gauge parts of these expressions exactly coincide with the ones presented in \cite{Ellis:1988tx}. ($N_5$ and $N_{10}$ used in \cite{Ellis:1988tx} are related to $N_h$ and $N_H$ used in this paper by the equations $N_5=2N_h$ and $N_{10}=2 N_H$.) As for the Yukawa parts, we did not manage to find them in earlier papers.

\end{document}